\documentclass[usenatbib]{mn2e}

\usepackage{graphicx,wasysym, color, soul, ulem, amssymb,epstopdf}

\def\rpd{\hbox{rad\,d$^{-1}$}}

\def\chisqr{\hbox{$\chi^2_{\rm r}$}}

\def\rstar{\hbox{$R_{\star}$}}

\def\sn{\hbox{S/N}}
\def\vrad{\hbox{$v_{\rm rad}$}}
\def\ms{\hbox{m\,s$^{-1}$}}

\def\kms{\hbox{km\,s$^{-1}$}}
\def\vsini{\hbox{$v \sin i$}}

\def\ptt{\hbox{$10^{-4} I_{\rm c}$}}

\def\degr{\hbox{$^\circ$}}

\def\omeq{\hbox{$\Omega_{\rm eq}$}}
\def\dom{\hbox{$d\Omega$}}

\begin{document}

\title[The evolving magnetic field of HD189733]{MOVES I. The evolving magnetic field of the planet-hosting star HD189733}

\makeatletter

\def\newauthor{%
  \end{author@tabular}\par
  \begin{author@tabular}[t]{@{}l@{}}}
\makeatother
\author[R.~Fares et al.]{\vspace{1.7mm} 
R.~Fares$^{1,2}$\thanks{E-mail:
rfares@oact.inaf.it},
V.~Bourrier$^3$, A.A.~Vidotto$^{3,4}$, C.~Moutou$^{5,6}$, M.M.~Jardine$^7$, P. Zarka$^8$\\
\vspace{1.7mm}
{\hspace{-1.5mm}\LARGE\rm
Ch.~Helling$^9$,  A.~Lecavelier~des~Etangs$^{10}$, J.~Llama$^{11}$, T.~Louden$^{12}$, P.J.~Wheatley$^{12}$}\\
\vspace{1.7mm}
{\hspace{-1.0mm}\LARGE\rm{D.~Ehrenreich$^3$}}\\
$^1$ INAF - Osservatorio Astrofisico di Catania, Via Santa Sofia 78, 95123 Catania, Italy  \\
$^2$ Department of Natural Sciences, School of Arts and Sciences, Lebanese American University, P.O. Box 36, Byblos, Lebanon\\
$^3$ Observatoire de l'Universit\'e de Gen\`eve, Chemin des Maillettes 51, Versoix, CH-1290, Switzerland \\
$^4$ School of Physics, Trinity College Dublin, Dublin-2, Ireland \\
$^5$ LAM--UMR 6110, CNRS \& Univ.\ de Provence, 38 rue Fr\'ederic Juliot-Curie, F--13013 Marseille, France \\
$^6$ CFHT, 65-1238 Mamalahoa Hwy., Kamuela, HI 96743, USA \\
$^7$ SUPA, School of Physics and Astronomy, Univ.\ of St~Andrews, St~Andrews, Scotland KY16 9SS, UK \\
$^8$ LESIA, Observatoire de Paris, CNRS, PSL/SU/UPMC/UPD/SPC, Place J. Janssen, 92195 Meudon, France \\
$^9$ Centre for Exoplanet Science, SUPA, School of Physics and Astronomy, Univ.\ of St~Andrews, St~Andrews, Scotland KY16 9SS, UK \\
$^{10}$ IAP-UMR7095, CNRS \& Universit\'e Pierre \& Marie Curie, 98bis boulevard Arago, 75014 Paris, France\\
$^{11}$ Lowell Observatory, 1400 W. Mars Hill Rd. Flagstaff. Arizona, 86001, USA\\
$^{12}$ Department of Physics, University of Warwick, Coventry CV4 7AL, UK\\
}

\date{}

\maketitle

\label{firstpage}

\begin{abstract}
HD189733 is an active K dwarf that is, with its transiting hot Jupiter, among the most studied exoplanetary systems. 
In this first paper of the Multiwavelength Observations of an eVaporating Exoplanet and its Star (MOVES) program, we present a 2-year monitoring of the large-scale magnetic field of HD189733.
The magnetic maps are reconstructed for five epochs of observations, namely June-July 2013, August 2013, September 2013, September 2014, and July 2015, using Zeeman-Doppler Imaging. We show that the field evolves along the five epochs, with mean values of the total magnetic field of 36, 41, 42, 32 and 37 G, respectively. All epochs show a toroidally-dominated field. Using previously published data of \cite{moutou07} and \cite{2010MNRAS.406..409F}, we are able to study the evolution of the magnetic field over 9 years, one of the longest monitoring campaign for a given star. While the field evolved during the observed epochs, no polarity switch of the poles was observed. We calculate the stellar magnetic field value at the position of the planet using the Potential Field Source Surface extrapolation technique. We show that the planetary magnetic environment is not homogeneous over the orbit, and that it varies between observing epochs, due to the evolution of the stellar magnetic field. This result underlines the importance of contemporaneous multi-wavelength observations to characterise exoplanetary systems. Our reconstructed maps are a crucial input for the interpretation and modelling of our MOVES multi-wavelength observations. 

\end{abstract}

\section{Introduction}\label{sec:intro}
Hot-Jupiters (HJs), i.e., gas giant exoplanets orbiting close ($\lesssim 0.1$~au) to their host stars, are useful laboratories to study the complex interactions (e.g. magnetospheric, tidal, ionisation) between exoplanets and their host-stars. Because of the short orbital distance of HJs, interactions are expected to be strong and could potentially be detected with current instrumentation.
The high stellar XUV fluxes that HJs are subjected to can lead to enhanced heating and atmospheric escape \citep[e.g.,][]{2007A&A...461.1185L,2009ApJ...693...23M,2009MNRAS.396.1012D,2011OLEB...41..503L,2012MNRAS.422.2024J,2013A&A...557A.124B,2013Icar..226.1678K}, which can be probed through transmission spectroscopy technique \citep{2003Natur.422..143V,2010ApJ...714L.222F,2010A&A...514A..72L,2012A&A...547A..18E,2012A&A...543L...4L,2013A&A...551A..63B}. In addition to the extreme radiation environment, HJs are also immersed in stellar winds, whose physical characteristics are unparalleled to those felt by the planets in our own solar system \citep{2005A&A...434.1191P,2007P&SS...55..618G,2009ApJ...703.1734V,2015MNRAS.449.4117V}. The extreme particle and magnetic environments of the stellar wind could lead to powerful reconnection events between the stellar and planetary magnetospheres \citep[e.g.][]{2004ApJ...602L..53I}, which may have the potential to enhance the stellar activity \citep{2000ApJ...533L.151C,2003ApJ...597.1092S,2008ApJ...676..628S,2013A&A...552A...7S}. In addition, as a result of the interaction between the planet's magnetosphere and the coronal material of the star, bow shocks might also form around HJs \citep[e.g.][]{2010ApJ...722L.168V,2011ApJ...738..166C,2013MNRAS.436.2179L,2015A&A...578A...6M}, explaining asymmetric UV transit features \citep{2010ApJ...714L.222F,2010ApJ...722L.168V}. Radio emission resultant from the interaction between the stellar magnetised wind and a magnetised exoplanet is expected to be several orders of magnitude larger than that of the largest emitter in our own solar system, Jupiter \citep{2007P&SS...55..598Z,2007A&A...475..359G,2008A&A...490..843J,2010MNRAS.406..409F,2012MNRAS.423.3285V,2015MNRAS.449.4117V,2015MNRAS.450.4323S}. A varying stellar magnetic field has also implication for the atmosphere of the orbiting planet leading to variability of the Cosmic Ray flux reaching the planetary atmosphere \citep{2013P&SS...77..152H,2013ApJ...774..108R}. Cosmic rays have been studied as one example for external ionisation sources that also effect the chemical compositions by possibly opening reaction paths to carbo-hydrate molecules \citep{2014ApJ...787L..25R,2014IJAsB..13..173R}.

HD189733 is an ideal system to study these different types of interactions. At a distance of $19.3$~pc, the bright (V=7.7) and active K2V star hosts a transiting HJ orbiting at $0.031\pm0.001$~au, i.e., at an orbital distance of $8.84\pm0.27~R_\star$ (see Table \ref{tab.stellar_properties}). The interactions within this system have been the subject of many studies in the literature  \citep[e.g.][]{2009MNRAS.395..335S,2010ApJ...722.1216P,2011ApJ...728L...6B,2011A&A...533A..50L,2013A&A...553A..52B,2013A&A...557A.124B,2013A&A...551A..63B,2013ApJ...773...62P,2015ApJ...814L..24L,2016ApJ...817..106B,2016MNRAS.459L.109B,2016MNRAS.462.1012B}. Because of its active host, HD189733b is in fast-changing radiation, particle and magnetic environments. Recent UV observations of the planetary transit showed that the properties of the hydrogen exosphere surrounding the planet are varying over time \citep{2012A&A...543L...4L}. An X-ray flare was detected 8h prior to the UV detection of atmospheric escape, suggesting that variations in the XUV emission of the host-star and/or its magnetised outflowing plasma  might have been the cause of the observed variability.

Variations in the quiescent stellar wind are also expected for this system. Through spectropolarimetric observations performed by \citet{moutou07} and \citet{2010MNRAS.406..409F}, it was shown that the large-scale magnetic field of HD189733 varied for the observed epochs. Recently, using the large-scale magnetic maps reconstructed by \citet{2010MNRAS.406..409F}, \citet{2013MNRAS.436.2179L} showed that the stellar wind of HD189733 presents inhomogeneities at the position of the planet, which could cause short timescale variations in the UV transit lightcurve (on the order of an orbital period), as well as longer timescale variations (on the order of a year, due to the change in the stellar magnetic field). These authors concluded that multi-wavelength data acquired simultaneously would provide the best tool for a comprehensive characterisation of the system. 
 
In this context, we started a multi-wavelength observational campaign of this system, in the frame of the MOVES collaboration (Multiwavelength Observations of an eVaporating Exoplanet and its Star, PI V. Bourrier). Observations of the star and the planet were obtained at similar epochs with ground-based and space-borne instruments in X-rays with Swift and XMM-Newton, UV with HST and XMM-Newton, optical spectropolarimetry with NARVAL and ESPaDOnS, and radio with LOFAR. In the present (first) paper of this collaboration, we use optical spectropolarimetric observations to reconstruct the surface magnetic field of HD189733 at five different epochs, contemporaneous to other sets of observations. These magnetic field maps will provide a crucial input to the analysis, modelling and interpretation of the multi-wavelength data sets that will follow in forthcoming papers. 
  
This paper is organised as follows. Section \ref{sec:obs} presents our observations. In Section \ref{sec:zdi}, we describe the magnetic imaging method we use. The results are shown in Section \ref{sec:results}, where we present the reconstruction of the magnetic field of HD189733 at five different epochs. We discuss the magnetic field evolution of HD189733 in Section \ref{sec:discussion}, based on this paper's results and results of \cite{moutou07} and \cite{2010MNRAS.406..409F}. Section \ref{sec:conclusions} presents the summary and conclusions of this work.

\section{Observations}\label{sec:obs} 

\begin{table}
\caption{Summary of the physical properties of HD189733 and its hot Jupiter, HD189733b. }
\label{tab.stellar_properties}
\begin{tabular}{lcl}
\hline
\hline
Physical property& value & reference \\
\hline
{\bf Star:} &&\\
Sp.~type & K2V & \\
V (mag)& 7.7 &\\
$T_{\rm eff}$ (K) & $5050 \pm 50$ & \citet{bouchy05}\\
$M_\star  (M_\odot$)  & $0.92 \pm 0.03$  &\citet{bouchy05} \\
$R_\star (R_\odot$)     & $0.76\pm0.01$ & \citet{winn07}\\
 \vsini  ~(\kms) & $2.97\pm0.22$ &  \citet{winn06} \\
$P_{\rm rot}$ (d) & $11.94\pm 0.16$ &  \citet{2010MNRAS.406..409F}\\  
$i_\star (\degr)$ & $\sim 85$ &\citet{2010MNRAS.406..409F}\\ 
$d\Omega~(\rm{rad~s}^{-1})$&$0.146\pm 0.049$& \citet{2010MNRAS.406..409F} \\ 
\hline
{\bf Planet:} &&\\
$i_{\rm orb} (\degr)$& $85.76\pm0.29$ & \citet{boisse09} \\
$M_{p} (M_{\jupiter})$&$1.13\pm0.03$& \citet{boisse09} \\
$R_p (R_{\jupiter})$&$1.154\pm0.032$& \citet{boisse09} \\
$P_{\rm orb} (d)$&$2.2185733\pm0.0000019$& \citet{boisse09} \\
$a$ (au)& $0.031\pm0.001$ & \citet{boisse09} \\
$i_{\rm spin-orbit} $& $0.85^{+0.32}_{-0.28}$&  \citet{triaud09} \\
\hline
\end{tabular}
\end{table}

Our spectropolarimetric data were obtained using both NARVAL spectropolarimeter at the TBL (2m) and ESPaDOnS at CFHT (3.6m). A spectropolarimeter is a spectrograph with a polarisation section, allowing measurement of the polarisation of the spectral lines. A circular polarisation spectrum is extracted from 4 subexposures taken each at a different angle of the polarisation waveplates. ESPaDOns and NARVAL are twin instruments, both operating in the optical domain (370 to 1000~nm) at a resolution of 65,000 in the polarisation mode. Data reduction is done automatically using Libre-Esprit, a fully automated reduction tool installed at TBL and CFHT \citep{donati97}. The spectra are normalised to a unit continuum, their wavelengths referring to the Heliocentric rest frame. Telluric lines are used to correct from spectral shifts due to instrumental effects. This correction secures a radial velocity (RV) precision of about 15~\ms~\citep{moutou07,morin08}.

Observations with both NARVAL and ESPaDOnS were obtained in service mode. Our NARVAL 2013 data (PI Bourrier) were collected  as follow: 5 spectra in May (11-13), 2 spectra in June (12-15), 5 spectra in July (01-04-05-08-14), 14 spectra in August (02 to 23), 13 spectra in September (02 to 24) and finally 5 spectra in October (08-09-12-13-18). ESPaDOnS' data were obtained via a filling program targeting planet-hosting stars (PI Moutou), 7 spectra were collected in September 2013. Table \ref{tab:logobs} presents the log of observations of all our data in 2013. In 2014, 15 spectra were collected between 01 September and 20 October. Another 18 spectra were collected between 20 May 2015 and 20 July 2015. The log of the 2014 and the 2015 (PI Bourrier) observations are presented in Table \ref{tab:logobs14}.

\begin{table*}

\caption{List of observations in 2013. The columns list, respectively, the dates of observation, Heliocentric Julian Date and UT time of observations (at mid-exposure), the peak S/N  (per 2.6 \kms\ velocity bin) of each observation
(around 700 nm), the rotational cycle of the star and orbital cycle of the planet calculated using the ephemeris of Eq.~\ref{eq:eph},  and the radial velocity (RV) of the star at each exposure, the rms noise level (relative to the unpolarized continuum level
$I_{\rm c}$ and per 1.8~\kms\ velocity bin) in the circular polarisation profile produced by LSD. The data were taken using NARVAL spectropolarimeter, except for 6 spectra collected using ESPaDOnS (marked by * next to the date of observation). The exposure time of each observation is 4$\times$900 s. Dates marked with \textdagger\ were not used for the mapping of the stellar field (see text for more details).}

\begin{tabular}{cccccccccc}
\hline
\hline
Date (UT)  & HJD  & UT &  \sn& Rot. Cycle & Orb. Cycle & \vrad &  $\sigma_{LSD}$ \\
(2013)&    (2,456,000+) & (h:m:s) &   &(239+) & (1297+)&(\kms)&(\ptt)  \\
\hline
12 May\textdagger  &  424.58158 & 01:55:55 &  630 & -6.0673 & -37.0955 & -2.134 & 0.50 \\
12 May\textdagger  &  424.62595 & 02:59:48 &  670 & -6.0636 & -37.0755 & -2.154 & 0.48 \\
13 May\textdagger  &  425.59553 & 02:15:52 &  680 & -5.9828 & -36.6385 & -2.340 & 0.46 \\
13 May\textdagger  &  425.63991 & 03:19:46 &  680 & -5.9791 & -36.6185 & -2.325 & 0.46 \\
14 May\textdagger  &  426.62177 & 02:53:31 &  660 & -5.8973 & -36.1759 & -2.029 & 0.47 \\
\hline
13 Jun   &  456.53057 & 00:38:37 & 440 & -3.4049 & -22.6948 & -2.385 & 0.81 \\
16 Jun   &  459.52985 & 00:37:17   & 540 & -3.1549 & -21.3429 & -2.045 & 0.62 \\
02 Jul   &  475.60489 & 02:24:02   & 650 & -1.8153 & -14.0973 & -2.075 & 0.51 \\
05 Jul   & 478.60456 & 02:23:22   & 650 & -1.5654 & -12.7452 & -2.477 & 0.47 \\
06 Jul   & 479.59405 & 02:08:10   & 580 & -1.4829 & -12.2992 & -2.045 & 0.56 \\
08 Jul   & 482.52265 & 24:25:11   & 690 & -1.2389 & -10.9792 & -2.218 & 0.47 \\
15 Jul   &  488.54927 & 01:03:14   & 380 & -0.7366 & -8.2627 & -1.906 & 0.92 \\
\hline
02 Aug  & 507.50215& 23:55:01 &   400& 0.8428 & 0.2801 & -2.298 & 0.87 \\
04 Aug  & 509.37413& 20:50:4&   660& 0.9988 & 1.1239 & -2.277 & 0.50 \\
05 Aug  & 510.51950& 24:20:01 &   440& 1.0942 & 1.6401 & -2.00& 0.71 \\
08 Aug  & 513.50839& 24:04:03 &   530& 1.3433 & 2.9873 & -2.124 & 0.63 \\
09 Aug  & 514.50144& 23:54:04 &   680& 1.426& 3.4349 & -2.241 & 0.46 \\
10 Aug  & 515.50391& 23:57:39 &   650& 1.5096 & 3.8868 & -2.016 & 0.49 \\
11 Aug  & 516.52369& 24:26:09 &   660& 1.5946 & 4.3464 & -2.294 & 0.49 \\
14 Aug  & 518.53989& 0:49:32 &   680& 1.7626 & 5.2552 & -2.326 & 0.46  \\
15 Aug  & 520.38524& 21:06:53 &   530& 1.9164 & 6.0870& -2.268 & 0.59  \\
18 Aug  & 523.37830& 20:56:59 &   630& 2.1658 & 7.4361 & -2.225 & 0.51  \\
19 Aug  & 524.46529& 23: 2:18 &   670& 2.2564 & 7.926& -2.058 & 0.47  \\
21 Aug  & 526.49842& 23:50:05 &   610& 2.4258 & 8.8425 & -2.027 & 0.53  \\
22 Aug  & 527.40235& 21:31:48 &   650& 2.5011 & 9.2499 & -2.37& 0.50  \\
23 Aug  & 528.50256& 23:56:09 &   640& 2.5928 & 9.7458 & -1.938 & 0.51 \\
\hline
02 Sep  &538.34180& 20:05:15 &   680& 3.4127 & 14.1807 & -2.352 & 0.47 \\
03 Sep  & 539.40804 & 21:40:42 &   660& 3.5016 & 14.6613 & -1.985 & 0.47 \\
04 Sep  & 540.37393 & 20:51:39 &   650& 3.5821 & 15.0967 & -2.287 & 0.51 \\
08 Sep  & 544.47392 & 23:15:58 &   610& 3.9237 & 16.9447 & -2.125 & 0.53 \\
10 Sep  & 546.42285 & 22: 2:36 &   670& 4.0862 & 17.8232 & -2.014 & 0.48 \\
11 Sep  & 547.39768 & 21:26:26 &   660& 4.1674 & 18.2626 & -2.387 & 0.48 \\
12 Sep  & 548.39552& 21:23:24 &   660& 4.2505 & 18.7123 & -2.082 & 0.30 \\
13 Sep*  & 548.86339& 08:37:12 &   1020& 4.2895 & 18.9232 &-2.082 & 0.30 \\
13 Sep  & 549.40705& 21:40:06 &   560& 4.3348 & 19.1683 & -2.346 & 0.60 \\
15 Sep  & 551.36003& 20:32:34 &   420& 4.4976 & 20.0486 & -2.211 & 0.76 \\
17 Sep*  & 552.89635& 09:25:02 &   900& 4.6256 & 20.7410&-1.986 & 0.37 \\
19 Sep  & 555.38839& 21:13:48 &   670& 4.8333 & 21.8643 & -2.242 & 0.34 \\
20 Sep*  & 555.73218& 05:28:55 &   960& 4.8619 & 22.0193 &-2.242 & 0.34 \\
20 Sep  & 556.39231& 21:19:33 &   640& 4.9169 & 22.3168 & -2.39& 0.49 \\
21 Sep  & 557.35420& 20:24:46 &   560& 4.9971 & 22.7504 & -2.183 & 0.34 \\
22 Sep*  & 557.88116& 09:03:4&   990& 5.0410& 22.9879 &-2.182 & 0.34 \\
24 Sep*\textdagger  & 559.89429& 09:22:47 &   870& 5.2088 & 23.8953 &-2.017 & 0.39 \\
24 Sep  & 560.32644& 19:45:07 &   670& 5.2448 & 24.0901 & -2.36& 0.33 \\
25 Sep*  & 560.80689& 07:17:02 &   970& 5.2848 & 24.3066 &-2.36& 0.33 \\
27 Sep*  & 562.89300& 09:21:16 &   880& 5.4587 & 25.2469 &-2.379 & 0.37 \\
\hline
08 Oct\textdagger   & 574.28920 & 18:53:07 &  690 & 6.4084 & 30.3837 & -2.332 & 0.46  \\
09 Oct\textdagger  &  575.28808 & 18:51:38 &  670 & 6.4916 & 30.8339 & -2.007 & 0.47  \\
12 Oct\textdagger   &  578.28726 & 18:50:50 &  610 & 6.7415 & 32.1857 & -2.329 & 0.53  \\
13 Oct\textdagger   & 579.29042 & 18:55:31 &  530 & 6.8251 & 32.6379 & -2.008 & 0.63  \\
18 Oct\textdagger   & 584.28323 & 18:45:50 & 560 & 7.2412 & 34.8884 & -2.037 & 0.60 \\
\hline
\end{tabular}
\label{tab:logobs}
\end{table*}

\begin{table*}
\caption{Same as Table \ref{tab:logobs} for the observations in 2014 and 2015. The exposure time is $4\times900$~s, expect for 28 May 2015 with an exposure time of $4\times800$~s, and 20 July 2015 where only two sub-exposures of 900~s each were taken. All spectra were obtained using NARVAL. Dates marked with \textdagger\ were not used for the mapping of the stellar field. }
\begin{tabular}{cccccccccc}
\hline
Date (UT)  & HJD  & UT &  \sn& Rot. Cycle & Orb. Cycle & \vrad &  $\sigma_{LSD}$ \\
 &    (2,456,000+) & (h:m:s) &   &(239+) & (1297+)&(\kms)&(\ptt)  \\
\hline
\hline
01 Sep 2014  & 902.40942 & 21:42:33 & 690 & 33.7517 & 178.2805 & -2.339 & 0.47 \\
02 Sep  2014 &  903.37339 & 20:50:43 &  660 & 33.8320 & 178.7150 & -1.966 & 0.49 \\
03 Sep 2014  &  904.37260 & 20:49:39 &  670 & 33.9153 & 179.1654 & -2.312 & 0.48 \\
05 Sep 2014  &  906.37382 & 20:51:33 &  680 & 34.0821 & 180.0674 & -2.193 & 0.46 \\
10 Sep 2014 &  911.37147 & 20:48:35 &  670 & 34.4985 & 182.3200 & -2.338 & 0.48 \\
11 Sep 2014 &  912.35651 & 20:27:07 &  640 & 34.5806 & 182.7640 & -1.948 & 0.51 \\
12 Sep 2014 &  913.39844 & 21:27:35 &  660 & 34.6675 & 183.2337 & -2.366 & 0.50 \\
24 Sep 2014\textdagger\  & 925.34813 & 20:16:19 &  90 & 35.6633 & 188.6199 & -2.041  & 5.20 \\
25 Sep 2014 &  926.34560 & 20:12:47 &  460 & 35.7464 & 189.0695 & -2.260 & 0.71 \\
26 Sep 2014 &  927.34206 & 20:07:48 &  600 & 35.8294 & 189.5186 & -2.105 & 0.55 \\
27 Sep 2014 &  928.33867 & 20:03:01 &  560 & 35.9125 & 189.9678 & -2.092 & 0.59 \\
\hline
15 Oct 2014\textdagger\ &  946.31976 & 19:37:60 &320 & 37.4109 & 198.0726 & -2.197 & 1.13 \\
19 Oct 2014\textdagger\ & 950.32639 & 19:48: 5 &  110 & 37.7448 & 199.8786 & -2.032 & 3.92 \\
20 Oct 2014\textdagger\ &  951.31730 & 19:35: 8 &  660 & 37.8274 & 200.3252 & -2.303 & 0.48 \\
26 Oct 2014\textdagger\ &  957.27473 & 18:34:38 & 600 & 38.3238 & 203.0105 & -2.219 & 0.56 \\

\hline
28 May 2015\textdagger\  & 1170.62294 & 02:53:29  & 550 & 56.1028 & 299.1750 & -2.392 & 0.61 \\
29 May 2015\textdagger\  & 1171.55663 & 01:17:53  & 530 & 56.1806 & 299.5958 & -2.077 & 0.64 \\
31 May 2015\textdagger\ &  3173.55224 & 01:11:19 & 590 & 56.3469 & 300.4953 & -2.193 & 0.56 \\
08 June 2015\textdagger\  & 1181.58663 & 01:59:55  & 510 & 57.0165 & 304.1168 & -2.331 & 0.64 \\
\hline
25 June 2015  & 1199.48188 & 23:27:23  & 600 & 58.5077 & 312.1829 & -2.459 & 0.56 \\
30 June 2015  & 1203.56315 & 01:24:06  & 560 & 58.8478 & 314.0225 & -2.207 & 0.64 \\
01 July 2015  & 1204.53004 & 00:36:21  & 570 & 58.9284 & 314.4583 & -2.259 & 0.61 \\
06 July 2015  & 3210.48886 & 23:36:40 &  560 & 59.4250 & 317.1442 & -2.351 & 0.59 \\
07 July 2015  &  3211.48835 & 23:35:53 &  600 & 59.5083 & 317.5947 & -2.059 & 0.55 \\
08 July 2015  & 3212.49073 & 23:39:15 &  580 & 59.5918 & 318.0465 & -2.241 & 0.53\\
09 July 2015  & 1213.47407 & 23:15:13  & 670 & 59.6738 & 318.4897 & -2.179 & 0.48 \\
10 July 2015  & 1214.49616 & 23:46:58  & 530 & 59.7589 & 318.9504 & -2.087 & 0.54 \\
12 July 2015  & 1215.52747 & 00:31:60  & 610 & 59.8449 & 319.4153 & -2.251 & 0.54 \\
12 July 2015  & 1216.51175 & 24:09:19  & 550 & 59.9269 & 319.8589 & -1.991 & 0.60 \\
13 July 2015  & 1217.52512 & 24:28:31 &  670 & 60.0113 & 320.3157 & -2.364 & 0.50 \\
14 July 2015 & 1218.52267 & 24:24:57 &  570 & 60.0945 & 320.7653 & -1.974 & 0.59 \\
16 July 2015  & 1219.52866 & 00:33:32  & 430 & 60.1783 & 321.2187 & -2.337 & 0.79 \\
20 July 2015  & 1224.42541 & 22:04:41 & 360 & 60.5864 & 323.4259 & -2.170 & 0.86 \\
\hline
\end{tabular}
\label{tab:logobs14}
\end{table*}

The equatorial rotation period of HD189733 is $\sim~12$~days \citep{2010MNRAS.406..409F}. For the sake of homogeneity with our previous studies of this star, we used the same ephemeris as in \citet{moutou07} and \citet{2010MNRAS.406..409F}. Rotational ($\rm{E_{\rm Rot}}$) and orbital ($\rm{E_{\rm Orb}}$) cycles are calculated by:
 
\begin{eqnarray}
T_{0}&=&\mbox{HJD~}2,453,629.389 + 12~\rm{E_{\rm Rot}}\nonumber\\
T_{0}&=&\mbox{HJD~}2,453,629.389 + 2.218575~\rm{E_{\rm Orb}} 
\label{eq:eph}
\end{eqnarray}

For cool stars, the polarisation signature in single lines is typically within the noise level.  Using the polarisation information of many spectral line simultaneously, we can extract the polarisation signal of the spectrum, this is known as multi-line technique Least-Square Deconvolution (LSD, \citealt{donati97}). LSD assumes that all lines have the same polarisation information, and extracts the polarisation signature by deconvolving the observed spectrum with a line mask. LSD calculates intensity (Stokes I) profiles, circular polarisation (Stokes V) profiles, as well as a null polarisation profiles (labelled N). These profiles are calculated using a combination of the sub-exposures taken at different angles of the waveplates. The Null profile is a check profile, it is calculated such to cancel out the stellar polarisation signature, and thus should contain no polarisation. It helps check for spurious or instrumental signatures. For more details on how these profiles are calculated, see \cite{donati97} and \cite{2017MNRAS.465.2734M}.
\newline
We compute a line mask for HD~189733 using Kurucz's lists of atomic line parameters (Kurucz CD-Rom 18) and a Kurucz model atmosphere with solar abundances, temperature set to $5000$~K and logarithmic gravity (in \hbox{cm\,s$^{-2}$}) set to 4.0. Only moderate-to-strong lines, featuring central depths larger than 40\% of the local continuum, are included in the mask (before any macroturbulent or rotational broadening); the strongest and broadest features (such as Balmer lines) are excluded. In the optical domain, this mask contains about 4000 lines. The LSD profiles calculated by deconvolving the stellar spectra with the mask have a S/N $\sim$ 30 times higher than the S/N in single lines with average magnetic sensitivity (see Tables \ref{tab:logobs} and \ref{tab:logobs14}), allowing for the detection of the polarisation signature.

We calculated the RV of each Stokes I profile by fitting a Gaussian profile to it. Since the star hosts a hot-Jupiter, the RV signatures varies over the planetary orbit. Our RV data agree (within the error bars) with the orbital solution of \citet{boisse09}. There is an offset between our values and theirs, $+0.06$~\kms\ for June-July 2013, $+0.13$~\kms\ for August 2013, $+0.10$~\kms\ for September 2013, $+0.13$~\kms\ for September 2014, and $+0.11$~\kms\ for June 2015. Such offsets are due to the use of different reduction pipelines (ESPaDOnS and Narval vs Sophie). For the tomographic Imaging, we shift each profile by its RV to centre all profiles around $0~\kms$.
\section{Model description and imaging method} \label{sec:zdi}
\label{model}

Magnetic fields, when present in the photosphere, cause splitting of the magnetically sensitive spectral lines via the Zeeman effect. The lines that form in such environments are polarised. Measuring the wavelength shift between spectral components of the line (when possible, which depends, among others, on the amplitude of this shift, the rotational broadening and the spectrograph resolution) allows us to calculate the longitudinal magnetic field of the star (see, e.g., \cite{2014A&A...563A..35S} for M dwarfs). The field topology (i.e. distribution, polarity, configuration), on the other hand, can not be determined by the wavelength shift alone, but is determined using a tomographic imaging technique, Zeeman-Doppler Imaging ZDI \citep{donati97}. Polarisation of the spectral lines depends on the orientation of the magnetic field relative to the line of sight \citep[see Fig 2 of][]{ansgarreview}. To map the field, spectra are collected during one or more stellar rotations. ZDI inverts these spectra/profiles into a magnetic topology that can produce the observed polarisation signatures. Because this problem is ill-posed, ZDI uses Maximum-Entropy regularisation to get the simplest magnetic map compatible with the data.

In the newest version of ZDI \citep{donati06}, the field is described by its poloidal and toroidal components \citep{Chand61}, with all the components expressed in terms of spherical harmonics expansion. The highest degree of spherical harmonic expansion used to map the field represents the ZDI map resolution around the equator. For slow rotators, such as HD189733, we limit the spherical harmonics to the lowest degrees ($l \leq 5$ , see \citealt{2010MNRAS.406..409F} for more details). ZDI follows an iterative procedure, it compares synthetic Stokes V profiles to the observed profiles collected during the stellar rotational cycles. In practice, the stellar surface is divided into 9000 grid cells of similar area. The contribution of each grid cell is then calculated in the weak field regime, and a synthetic Stokes profile for each observed rotation phase is produced.

In addition to the field intensity and distribution, ZDI also gives an indication on the stellar inclination (up to $\sim 10$\degr\ accuracy), on the stellar rotation period \citep[see, e.g.][]{alvarado15}, and on the differential rotation (DR). We describe DR using a solar-like DR law, where the rotation rate at a latitude  $\theta$ is defined by $\Omega(\theta) = \omeq - \dom  \sin^2(\theta)$, where $\omeq$ and $\dom$ are respectively the rotation rate at the equator, and the difference in rotation rate between the pole and the equator. In practice, to measure DR, we reconstruct the magnetic image at a given magnetic energy for a pair of fixed (\omeq, \dom), and obtain the reduced-chi squared (\chisqr) of the fit. We investigate the \chisqr\ values of the 2D parameter space of (\omeq, \dom). 
The optimum DR parameters are the ones minimizing \chisqr. They are obtained by fitting the surface of the \chisqr\ map with a paraboloid around the minimum value of \chisqr.

The Null profiles, for each epoch, are used as a test to check for spurious polarisation. In practice, we fit these profiles with a zero magnetic field configuration. Since the Null profiles should contain no polarisation, the \chisqr\ of the fit should be one or less. A \chisqr\ greater than one indicates either a spurious polarisation signature, or an underestimation of the error bars. If a systematic signature is found in the profiles, it indicates a spurious origin of the signature. We calculate a mean signature of the Null profiles, and subtract it from the Stokes V profiles to eliminate this spurious feature.

\section{Results}
\label{sec:results}

ZDI requires the use of data covering one or more stellar rotations to derive the photospheric magnetic map. In some field configurations \cite[see, e.g.][]{2008MNRAS.390..567M} the large-scale magnetic field of the star is stable for many stellar rotations, and also over many years. In these cases, one can combine data collected over many rotations together as one dataset. To estimate whether the field is stable over many stellar rotations, we compare the quality of the fit and the shape of the polarisation profiles at the same rotational phases. 
\newline
Our observations cover many stellar rotations at each observing epoch, from June 2013 to July 2015 (see Tables \ref{tab:logobs} and \ref{tab:logobs14}). 
\newline
We performed a series of tests on the data, reconstructing the maps using a combination of datasets for each epoch. We find that using Stokes V profiles spread over more than two consecutive stellar rotations to reconstruct the map worsens the quality of the fit. We therefore adopted datasets of up to two stellar rotations for each reconstructed map. Some spectra were not used in the reconstruction because they were collected with a rotational phase gap of more than a stellar rotation in respect to the main dataset (see Tables \ref{tab:logobs} and \ref{tab:logobs14}).

\subsection{Differential rotation}

Differential rotation distorts the magnetic regions on the stellar surface. Spectropolarimetric data can therefore be used to estimate the level of DR.
\newline
We applied the same technique as in \cite{2010MNRAS.406..409F}  to search for differential rotation in our data (explained in Section \ref{model}). In this study, we were able to detect DR for August 2013 dataset. The \chisqr\ map in the  (\omeq, \dom) space is shown in Fig \ref{fig:DRaug}. A well defined \chisqr\ minimum is found for $\dom = 0.11\pm 0.05~\rpd$ and \omeq\ of $0.535 \pm 0.004~\rpd$, corresponding to a rotation period at the equator  $11.7\pm0.1$d. 
The shear value \dom\ is consistent with the one measured by \cite{2010MNRAS.406..409F}  at $\dom = 0.146 \pm 0.049~\rpd$. These measurements, within their uncertainties, do not reveal a variation of DR. 
\newline
HD~189733 is a slow rotator for which the Fourier Transform of the Intensity profile technique presented by \cite{2002A&A...384..155R} can not be applied. There are no measurements of the DR of this star in the literature, apart from the recent work of  \cite{cegla16}. \cite{cegla16} modelled the Rossiter-McLaughlin effect to probe planetary parameters and stellar differential rotation, and found a $\dom\ \textgreater\ 0.12~\rpd$ for HD~189733, in agreement with our findings.
\newline
Stellar differential rotation is not constant across spectral types and rotation rates. Many observational and theoretical studies have addressed the question of its variation \cite[see, e.g., ][]{2005MNRAS.357L...1B,2006A&A...446..267R,2007AN....328.1030C,2012ApJ...756..169A,2013A&A...560A...4R,2016A&A...591A..43D,2016MNRAS.461..497B}. Using Kepler photometric data,  \cite{2013A&A...560A...4R} and \cite{2016MNRAS.461..497B} studied the DR for a large sample of stars. HD~189733 DR is within the range of the shear values of Kepler stars both as a function of temperature and as a function of rotation. \cite{2016MNRAS.461..497B} propose an empirical relation between the shear value, the effective temperature of the star and its rotation. For HD~189733, the predicted value of the \dom\ is $0.076^{+0.004}_{-0.001}$. Our value is slightly higher, their value is, however, within the error bars of our measurement.

\begin{figure}
\includegraphics[scale=0.35, angle=-90]{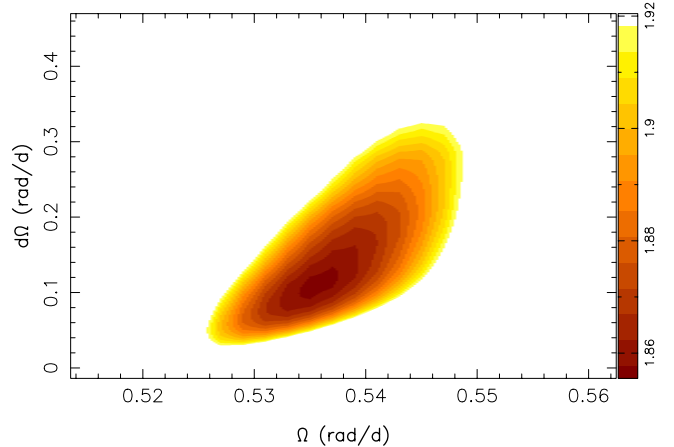}
\caption[]{Variations of \chisqr~as a function of \omeq~and \dom~, derived from the modelling of the Stokes V data set for August 2013. The outer colour contour corresponds to a 3.5\%~increase in the \chisqr~, and traces a 3 $\sigma$ interval for both parameters taken as a pair.}
\label{fig:DRaug}
\end{figure}

\subsection{Magnetic maps}
\label{sec:maps}

We reconstruct the magnetic maps for five observing epochs. The observed Stokes V (in red) and a 1-sigma error-bar are shown in Fig \ref{fig:profils}, along with the ZDI reconstructed circular polarisation (in black). The observed profiles show a detection of the polarisation signature (false-alarm probability less than $10^{-5}$, see \citet{donati97} for the classification of the detections). The reconstructed maps are shown in Fig \ref{fig:maps}. As mentioned previously, each dataset used to reconstruct a magnetic map consisted of no more than two stellar rotations.

\begin{itemize}
\item June/July 2013

\vspace{0.1cm}

We reconstructed the map of June/July 2013, although our phase coverage is not optimal. The \chisqr\ of the reconstruction is 1.15. The mean magnetic field is of 36~G. 61\% of the magnetic energy is in the toroidal component. For the poloidal component, we define axisymetric modes as having $m=0$ and $m \textless l/2$ ($m$ and $l$ being the order and the degree of the associated Legendre polynomial, used to describe the field as spherical harmonics expansion). About 40\% of the poloidal field is axisymetric. 
\newline

\item August 2013
\vspace{0.1cm}

For August 2013, the data samples the stellar rotation well. 
However, the Null profile shows a systematic signature in the core of the profile. Fitting these Null profiles with no magnetic field configuration leads to a \chisqr\ of 1.3. We calculated a mean profile for all the Null profiles. Subtracting the mean Null profile from the August 2013 Null spectra leads to a \chisqr\ of 0.7 when fitting the corrected Null profiles with no magnetic configuration. Given that the signatures in the Null profiles show a systematic trend, it is more likely to be a spurious signature rather than an underestimation of the error bars. We subtract the mean Null profile signature of the Stokes V profiles. The map is reconstructed with a \chisqr\ of $1.85$. The  average surface magnetic field is $40~\rm{G}$. 50\% of the total energy is in the toroidal component of the field. $2\%$ of the poloidal energy is in the axisymetric modes, and mainly in modes with $m=0$. The poloidal component is thus mainly non-axisymetric. Spherical harmonics modes with $l \textgreater 3$ (i.e. modes higher than the dipole, quadrupole and octupole) contribute to $\sim$ 38\% of the poloidal energy. 
\newline

\item September 2013
\vspace{0.1cm}

In September 2013, we have data from ESPaDOnS and NARVAL, covering two stellar rotations. We reconstructed a map using those data with a \chisqr\ of $2.15$. We notice a systematic difference in the Intensity profile depth between Narval's spectra and ESPaDOnS' spectra. The difference in depth is not significant enough to require different modelling of the intensity profiles between instruments. This difference is due to the normalisation of the spectra at the telescope. 

The average surface magnetic field is $42~\rm{G}$. 59\% of the energy in the toroidal component. The poloidal field is still mainly non-axisymetric (88\%, see Table \ref{table:stat}). About 50\% of the poloidal component is in the octupolar mode, higher orders contribute to $\sim 45\%$.
 \newline
 \item September 2014
\vspace{0.1cm}

In September 2014, the average magnetic field drops to $31~\rm{G}$ (see Fig \ref{fig:maps}). The \chisqr\ of the fit is $1.25$. The field remains mainly toroidal, with $78\%$ of the total energy stored in the toroidal components. The poloidal field is strongly non-axisymetric, and only $12\%$ of the energy lies within axisymmetric modes. 
\newline

\item July 2015
\vspace{0.1cm}

In July 2015, the magnetic field, and in particular the radial component, has changed significantly relative to September 2014. The radial component changes polarity around the equator between September 2014 and July 2015. The poloidal component contributes to 15\%~of the total energy. The average magnetic field is of 37~G.
\end{itemize}
The characteristics of each reconstructed map are listed in Table \ref{table:stat}. ZDI does not provide error bars for the reconstructed map. However, statistical errors can be calculated by varying the input parameters (e.g., \vsini, stellar inclination, \omeq, \dom) within their error bars. Error bars on field characteristics are calculated by comparing the characteristics of the reconstructed field for the best set of input parameters to the characteristics of maps reconstructed by varying those best input parameters within their error bars. We follow \cite{2016MNRAS.459.4325M} for error bar calculations: we fix (\omeq,\dom) and vary \vsini\ within its error bars (2.75 \kms\ to 3.2 \kms , Table \ref{tab.stellar_properties}). We then do a new set of maps fixing \vsini\ and varying (\omeq, \dom) within their error bars. Error bars in table \ref{table:stat} represent the highest values of our procedure.

\begin{figure*}

\center{\hbox{\includegraphics[scale=0.55,angle=-90]{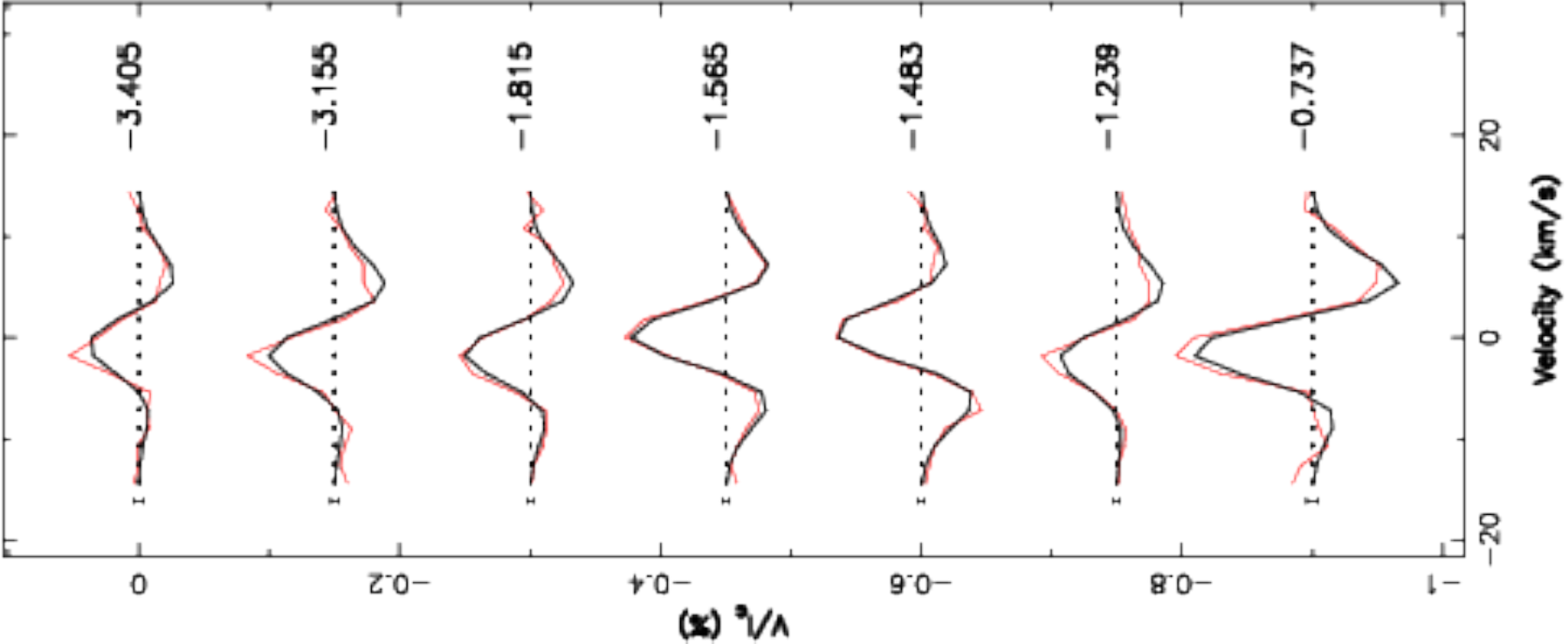}
\includegraphics[scale=0.55,angle=-90]{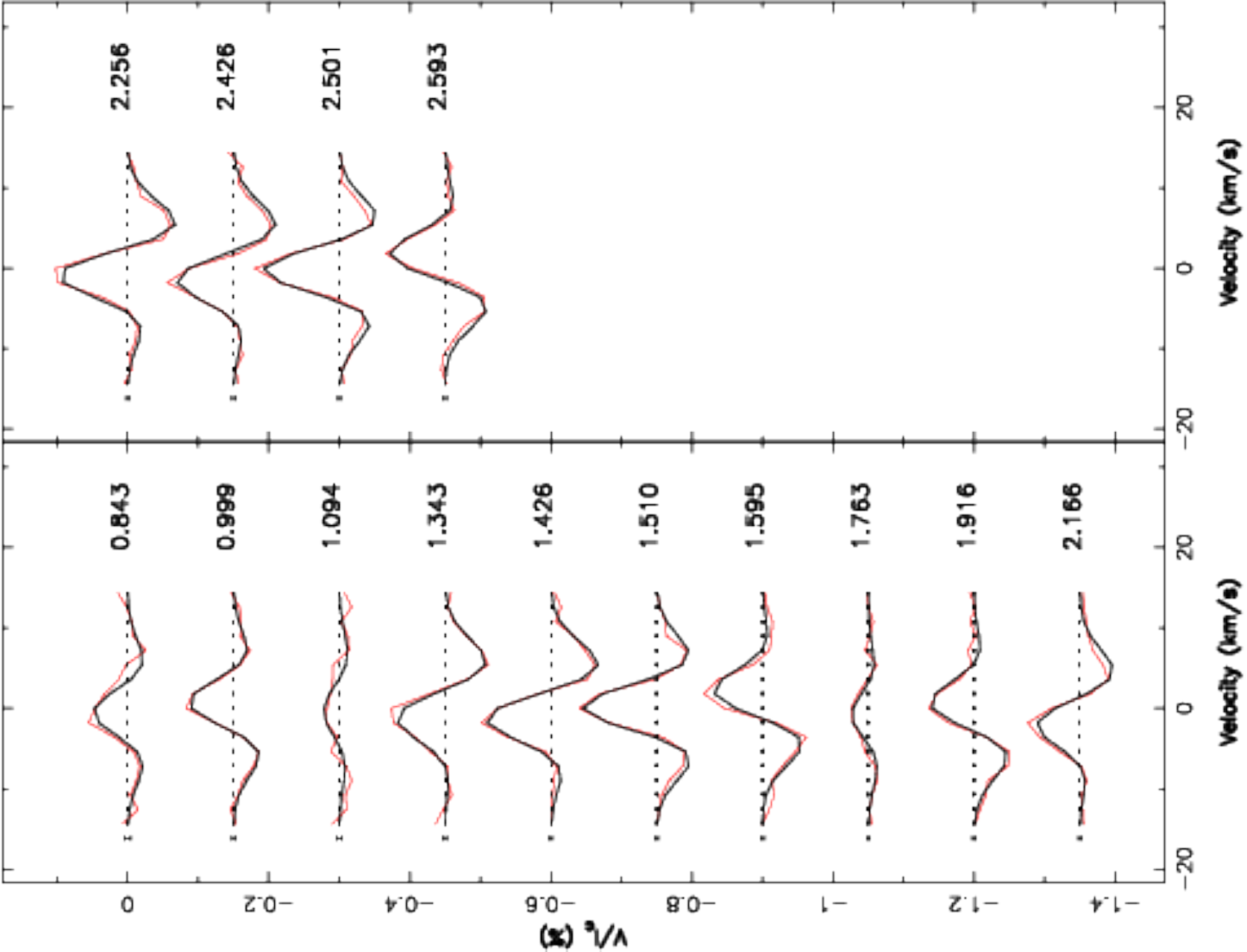}
\includegraphics[scale=0.55,angle=-90]{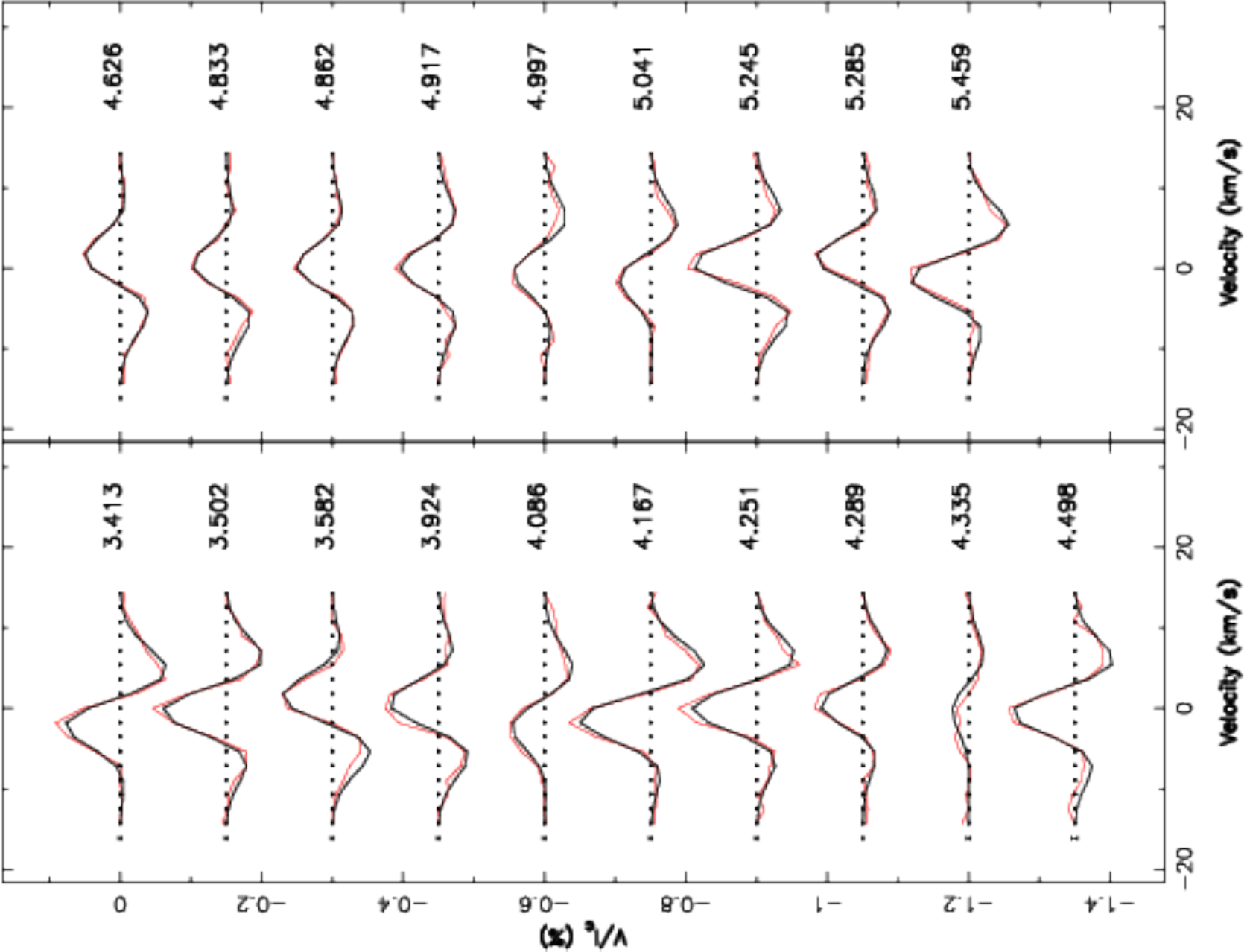}}
\hbox{\hspace{20mm}\includegraphics[scale=0.55,angle=-90]{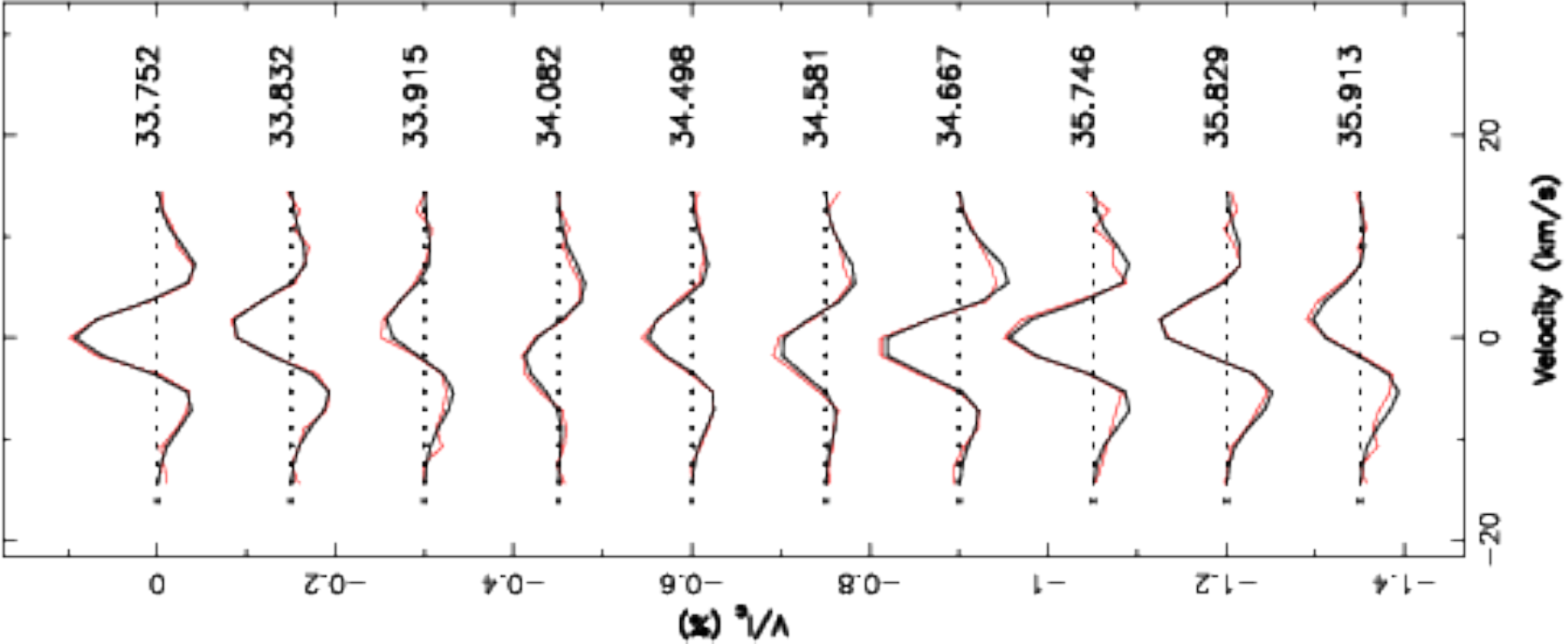}\hspace{25mm}
\includegraphics[scale=0.55,angle=-90]{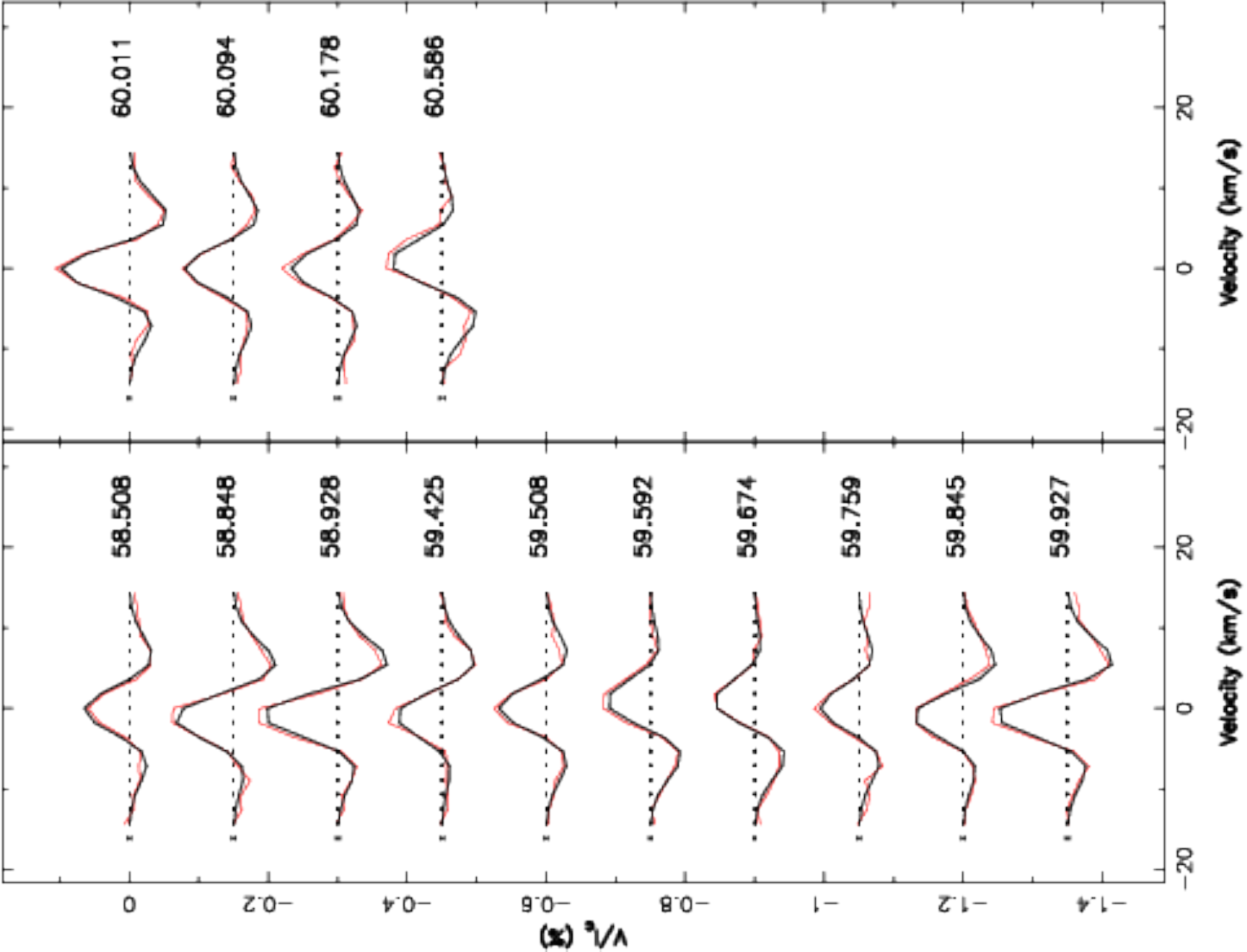}}}
\caption{ Observed Stokes V of HD189733 in red lines, and the fitted Stokes V profiles using ZDI in black, for June-July 2013 (top left), August 2013 (top middle), September 2013 (top right), September 2014 (bottom left) and July 2015 (bottom right).  The rotational cycle of each observation (as listed in Table \ref{tab:logobs}) and 1$\sigma$ error bars are also shown next to each profile.  }
\label{fig:profils}
\end{figure*}

\begin{figure*}
\center{\hbox{\hspace{12mm}\includegraphics[scale=0.65]{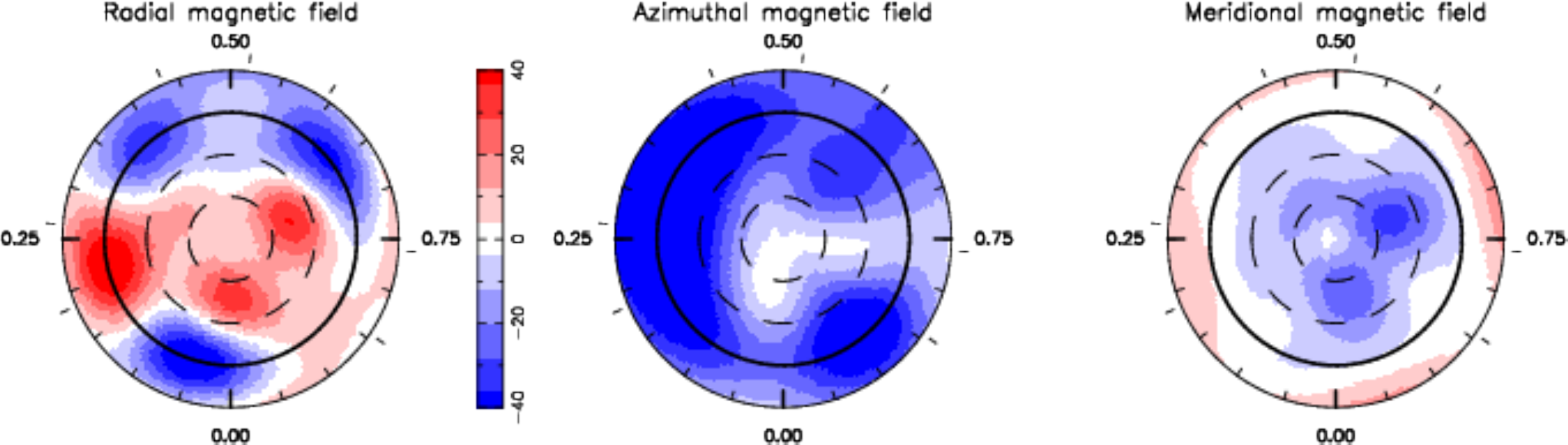}\hspace{3mm}}
\hbox{\hspace{12mm}\includegraphics[scale=0.65]{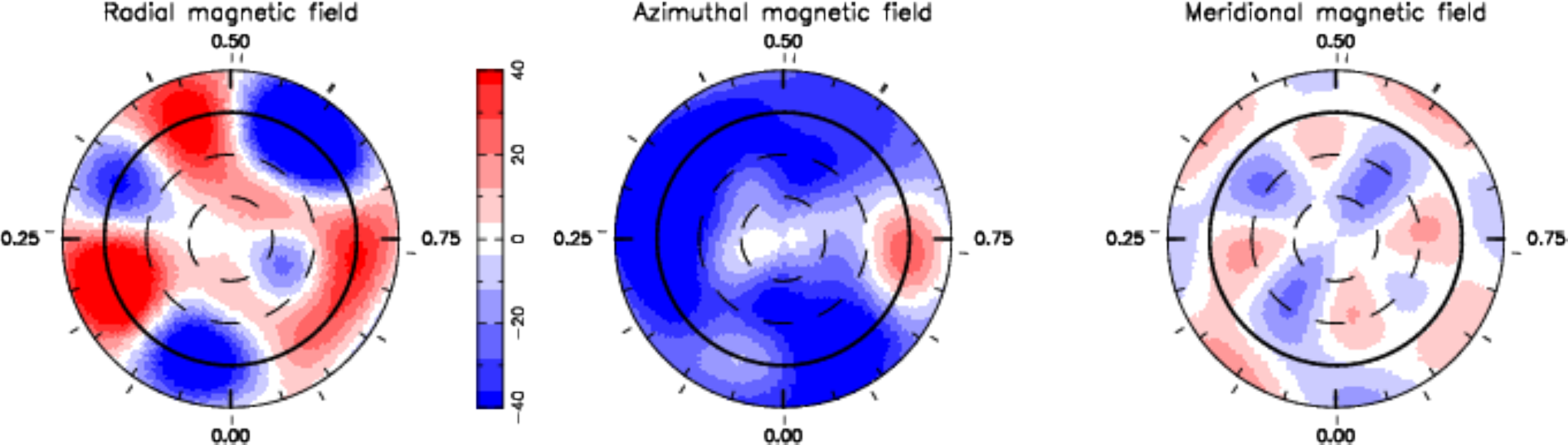}\hspace{3mm}}
\hbox{\hspace{12mm}\includegraphics[scale=0.65]{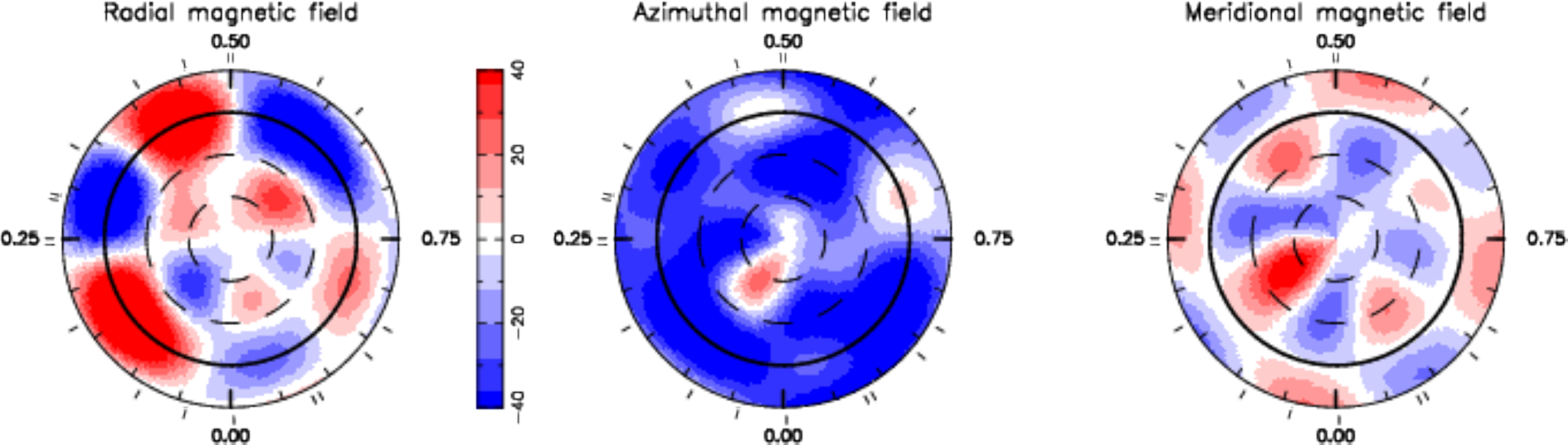}\hspace{3mm}}
\hbox{\hspace{12mm}\includegraphics[scale=0.65]{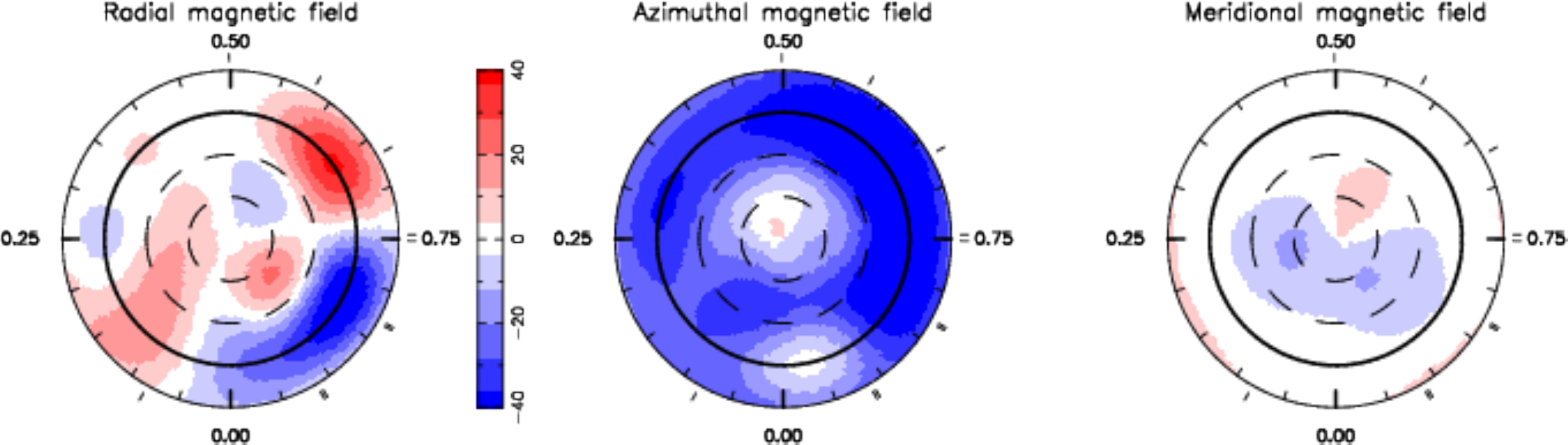}\hspace{3mm}}
\hbox{\hspace{12mm}\includegraphics[scale=0.65]{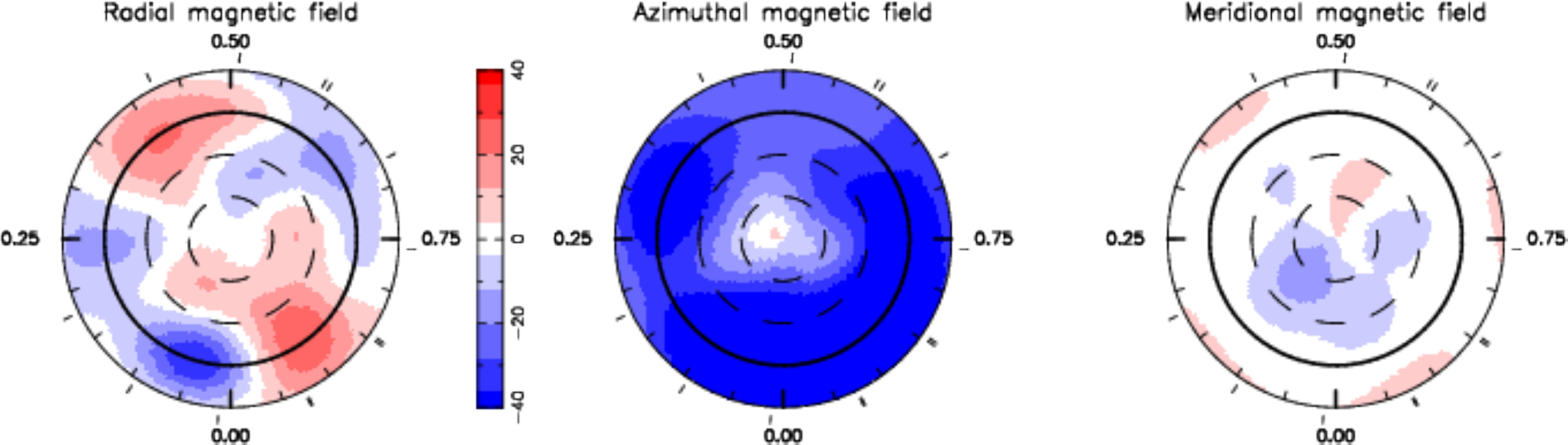}\hspace{3mm}}
}

\caption{The reconstructed maps of HD~189733 for June-July 2013 (top row) August 2013 (second row), September 2013 (third row), September 2014 (fourth raw) and July 2015 (bottom row). The maps are in a polar flattened projection down to latitudes of $-30\degr$, the equator is represented by the bold circle. The radial, azimuthal and meridional field components are shown. The magnetic flux values are labelled in G. Radial ticks around each map indicate the rotational phase of our observations.}
\label{fig:maps}
\end{figure*}

\subsection{Extrapolation of the magnetic field to investigate the corona and the planetary orbit}
\label{sec:extrapol}

The interactions between the stellar wind and the planetary magnetosphere might trigger planetary emission, such as radio emission or bow shock formation \citep[see, e.g.][]{2010MNRAS.406..409F,2013MNRAS.436.2179L}, and it might influence the ionisation state of the planetary atmosphere \citep{2013ApJ...774..108R,2014IJAsB..13..173R}. In this Section, we examine the stellar magnetic field in the corona up to the planetary orbit using a potential field extrapolation of the surface magnetic field. For this purpose, we use the Potential Field Source surface (PFSS) code of  \citealt[][]{jardine02}, originally developed for the Sun \citep{1998ApJ...509..435V} based on \cite{1969SoPh....9..131A}. The potential field extrapolation assumes that there is no electric current in the corona. The components of magnetic field in the corona are described using a spherical harmonics decomposition. This technique delivers satisfactory results when compared to wind modelling of the solar corona \citep{2006ApJ...653.1510R}.

The PFSS model extrapolates the magnetic field considering two boundary conditions: the first one being that the field is purely radial at a surface called the Source Surface, the second boundary condition is the observed field geometry at the surface of the star. We assume that the Source Surface is at 3.4~\rstar. Our radial magnetic maps are used as a boundary condition at the surface of the star, thus giving a realistic model of the radial field. 

The extrapolated field in the stellar corona for the five maps presented in this paper are shown in Fig \ref{fig:extrapol}. This figure shows how different surface field configurations produce different field line configuration in the corona. Since the magnetic field lines in the corona are not those of a very simple configuration, and since the planet and the star are not synchronised, the planet crosses different field configuration on its orbit, as well as from one orbit to the other (the planet crosses in front of the same stellar field configuration after one beat period (of rotation and orbital periods), rather than after one orbital period). We calculate, for each observing epoch, the footprints of field lines connecting the stellar surface to the position of the planet on its orbit . They are shown in Fig \ref{fig:footprints}.

\begin{figure*}
\center{
\hbox{\includegraphics[scale=0.3]{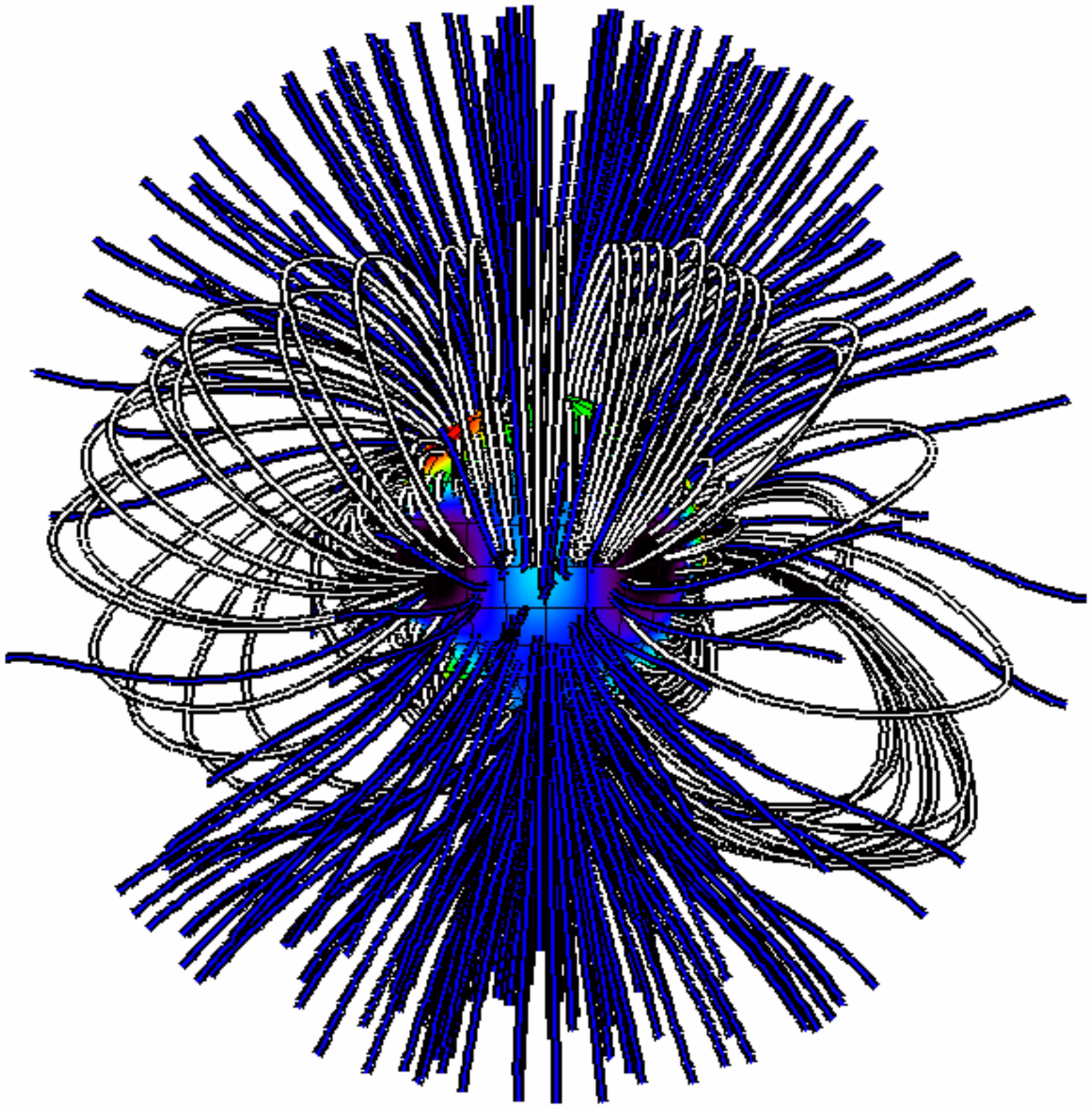}\hspace{2mm}\includegraphics[scale=0.3]{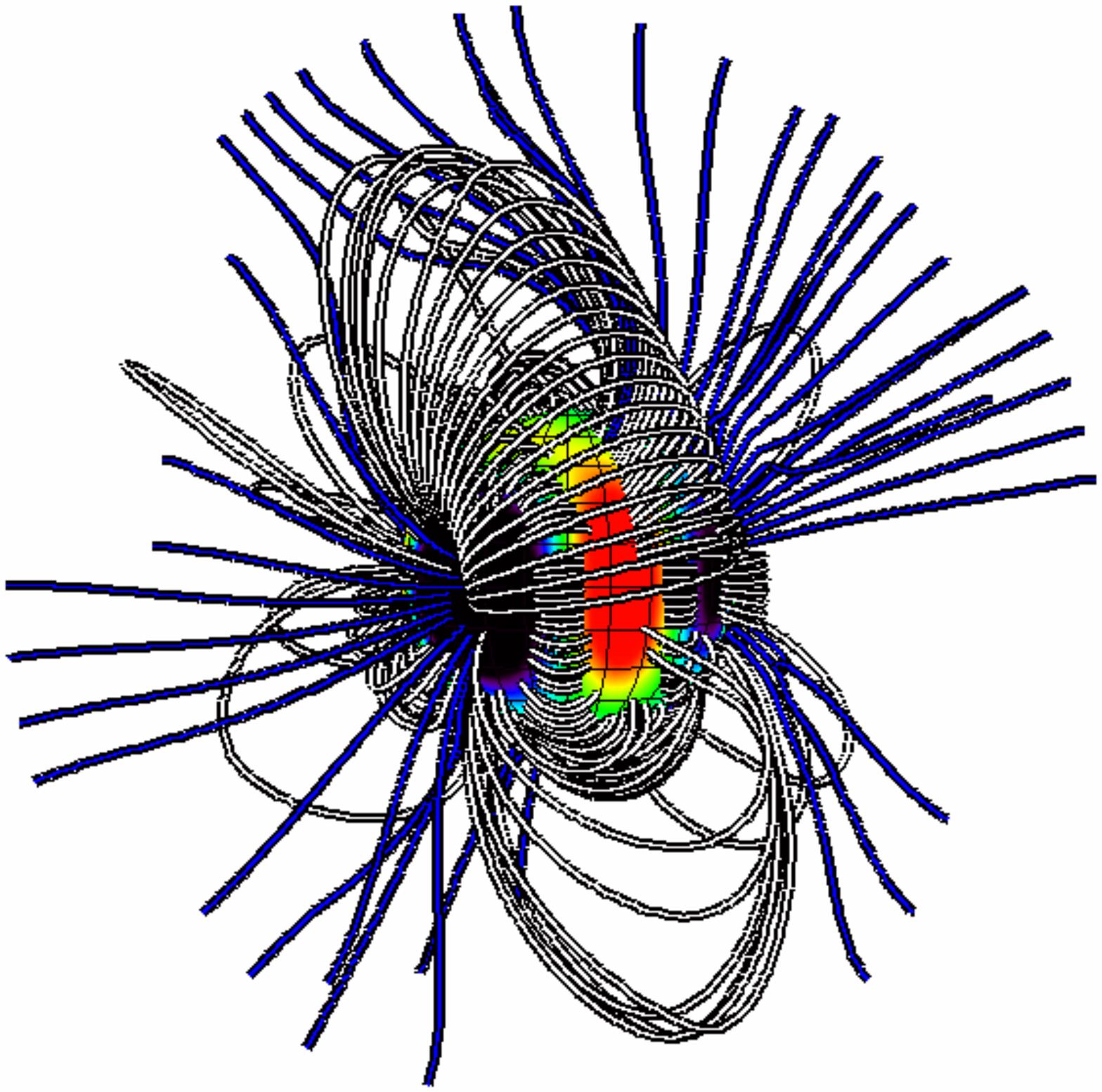}\hspace{2mm}\includegraphics[scale=0.3]{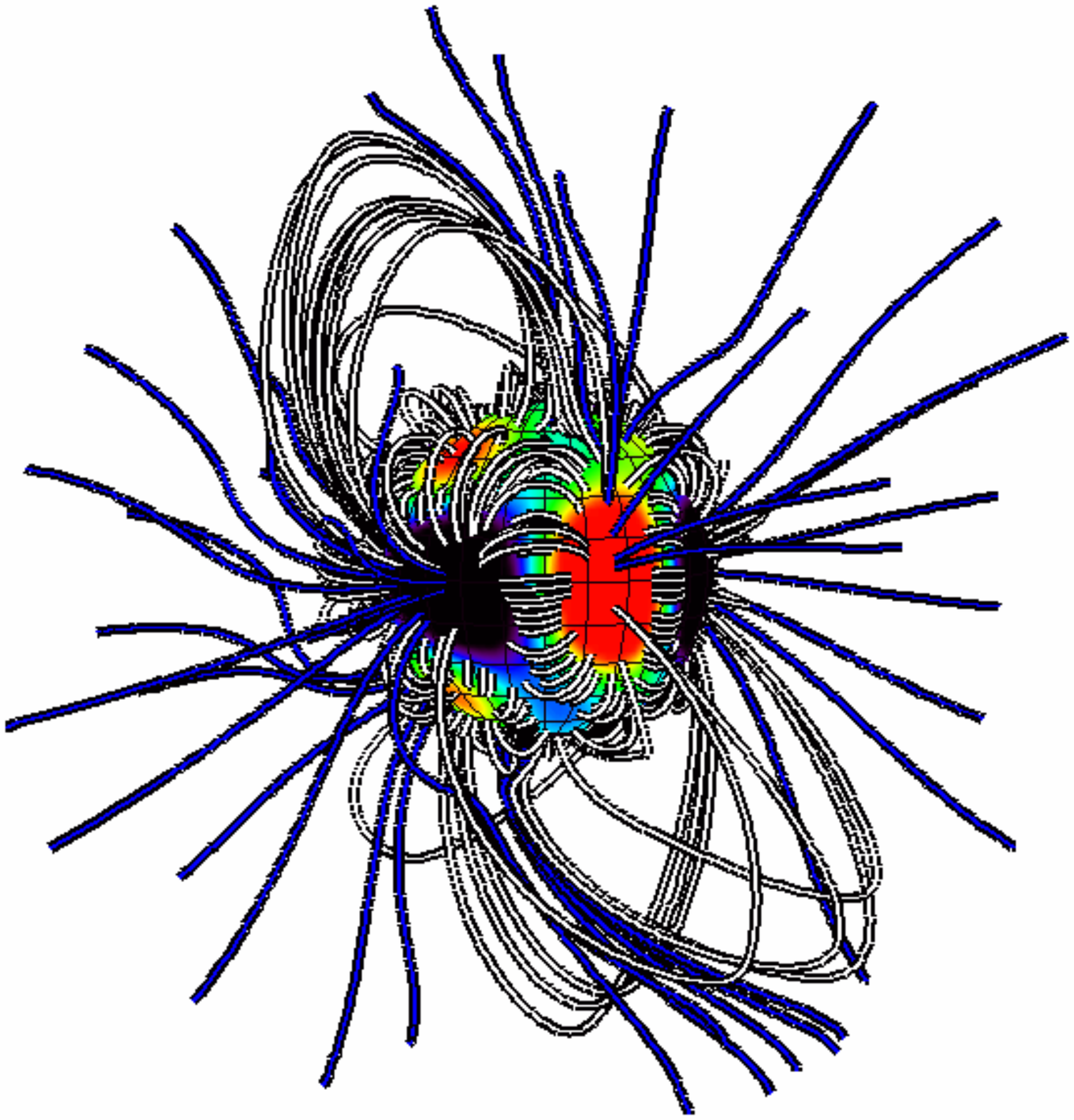}}
\hbox{\hspace{12mm}\includegraphics[scale=0.3]{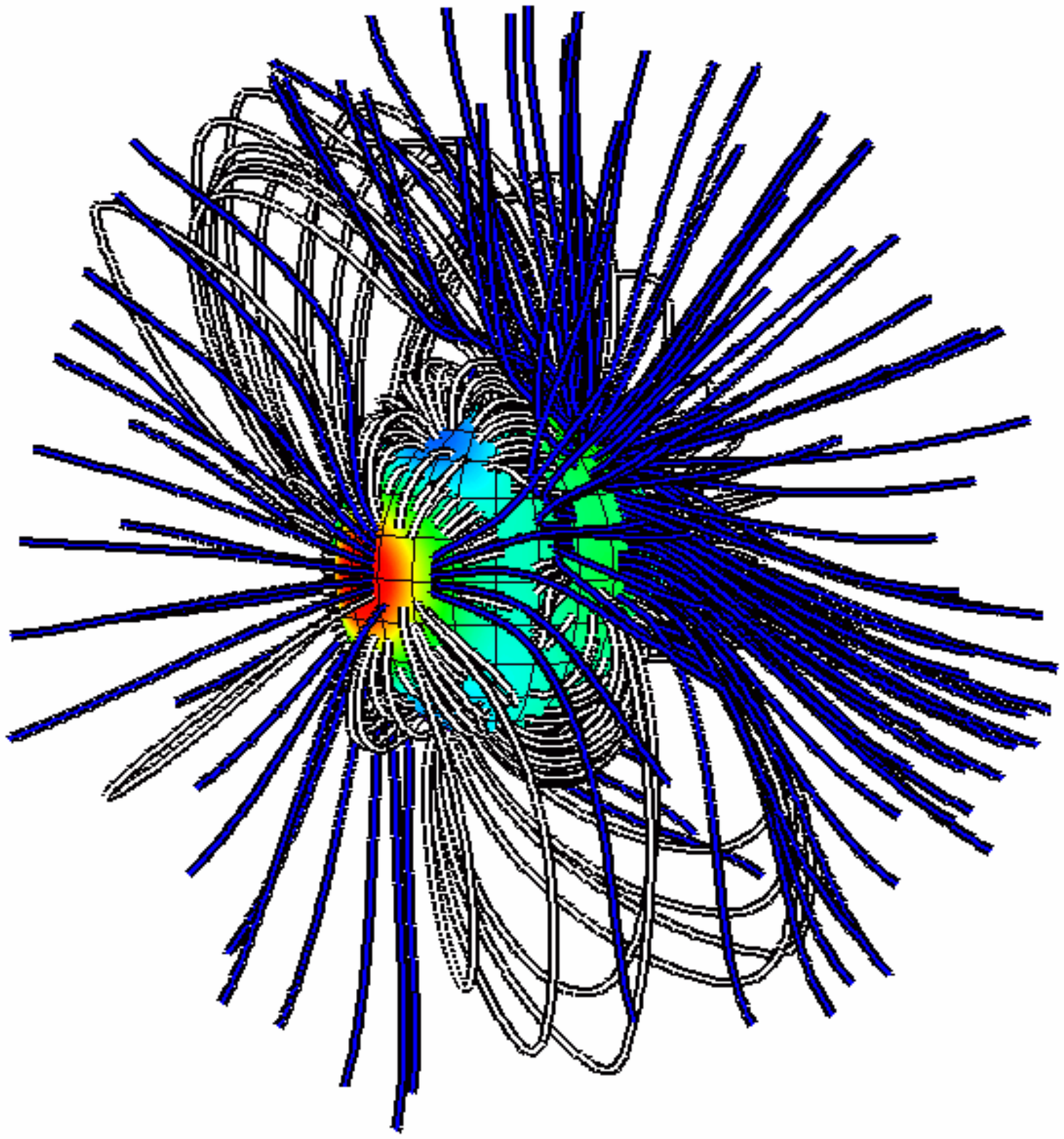}\hspace{12mm}\includegraphics[scale=0.3]{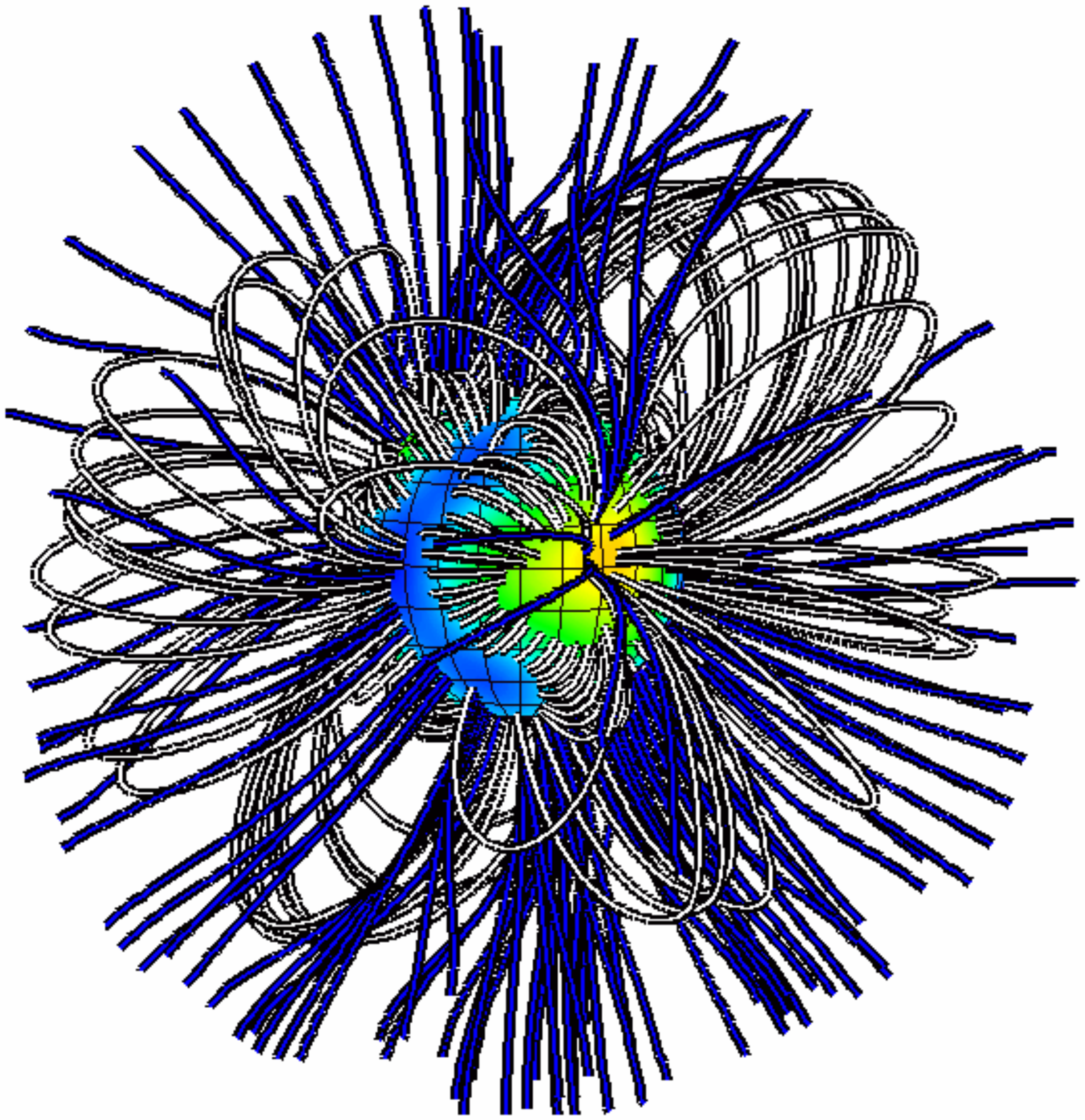}}}

\caption{The extrapolated magnetic field of HD~189733 for June-July 2013 (top left), August 2013 (top middle), September 2013 (top right), September 2014 (bottom left) and July 2015 (bottom right). White lines correspond to the closed magnetic lines, blue ones to the open field lines (reaching the source surface). The star is shown at the same rotational phase (0.5) to better visualise the differences in magnetic field topology at each observing epoch. The star is viewed almost equator on ($\sim 5$\degr), the inclination of the system on the sky (as seen from Earth). }

\label{fig:extrapol}
\end{figure*}

\begin{figure*}
\center{
\hbox{\hspace{4mm}\includegraphics[scale=1.5]{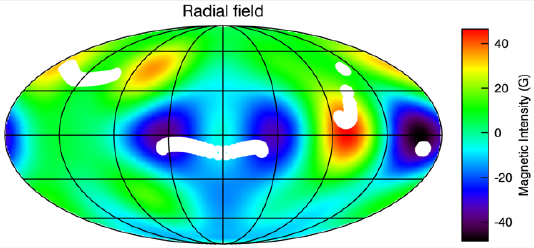}\hspace{5mm}\includegraphics[scale=1.5]{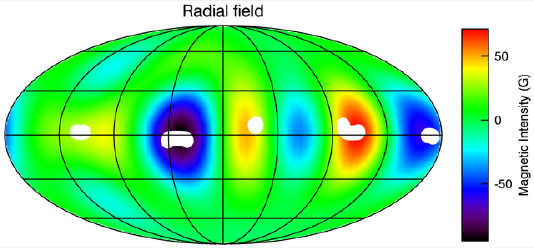}}
\hbox{\hspace{4mm}\includegraphics[scale=1.5]{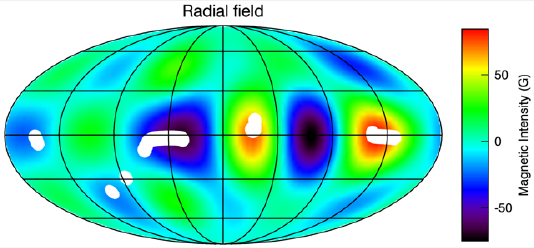}\hspace{5mm}\includegraphics[scale=1.5]{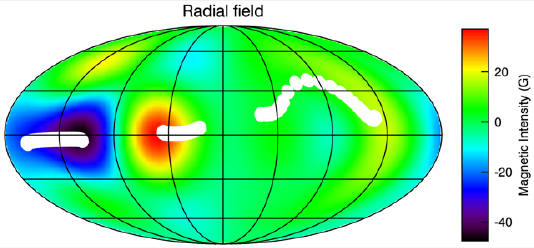}}
\hbox{\hspace{4mm}\includegraphics[scale=1.5]{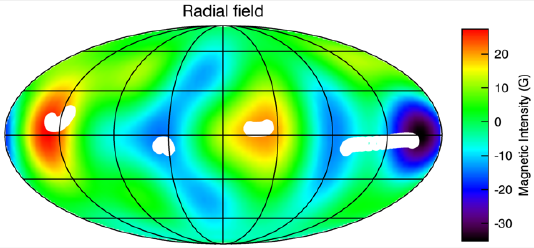}}}

\caption{The radial magnetic field of HD~189733 for June-July 2013 (top left), August 2013 (top right), September 2013 (middle left), September 2014 (middle right) and July 2015 (bottom). White dots represent the footprints of the field lines connecting the stellar surface to the position of the planet on its orbit. }

\label{fig:footprints}
\end{figure*}

The PFSS model allows the calculation of the field value and orientation at each position in the corona, out to the Source Surface. Since the planet is outside the Source Surface, we proceed as follows to calculate the field at its position. First, we calculate the position of the sub-planetary point on the Source Surface, we then calculate the energy budget at this point. We remind the reader here that at the Source Surface, the stellar field is purely radial, the meridional and azimuthal components are negligible. Assuming that the magnetic flux is conserved over spherical shells from the source surface to the planetary orbit, we calculate the decay of the magnetic field between the Source Surface and the orbit. In Fig \ref{fig:Bplanet}, we plot the field value at the planetary orbit for each epoch of observations (including June 2007 and July 2008). The maximum value the field reaches at the planetary orbit can change by 100\% between different epochs, which supports the importance of simultaneous observations when studying Star-Planet Interactions. We investigate the effect the error bars on the maps could have on the calculated magnetic field at the planetary orbit. To do so, for each epoch, we use as boundary condition each of the maps calculated for a range of \vsini, \omeq, and \dom\ (see Section \ref{sec:maps}), extrapolate the magnetic field for each map, and calculate the magnetic field at the position of the planet. We find that the mean difference between the field values presented in this paper and those calculated for the the maps reconstructed for sets of \vsini, \omeq, and \dom\ is of the order of 3mG. The difference can reach up to 70\% of the mean value over a small fraction of the orbit (for one epoch, all other epochs showed a smaller maximum difference value). This shows the robustness of our results: at the position of the planet, the variation of the stellar magnetic field is real and not affected by the error-bars on the maps.

\begin{figure*}
\center{\includegraphics[scale=0.5]{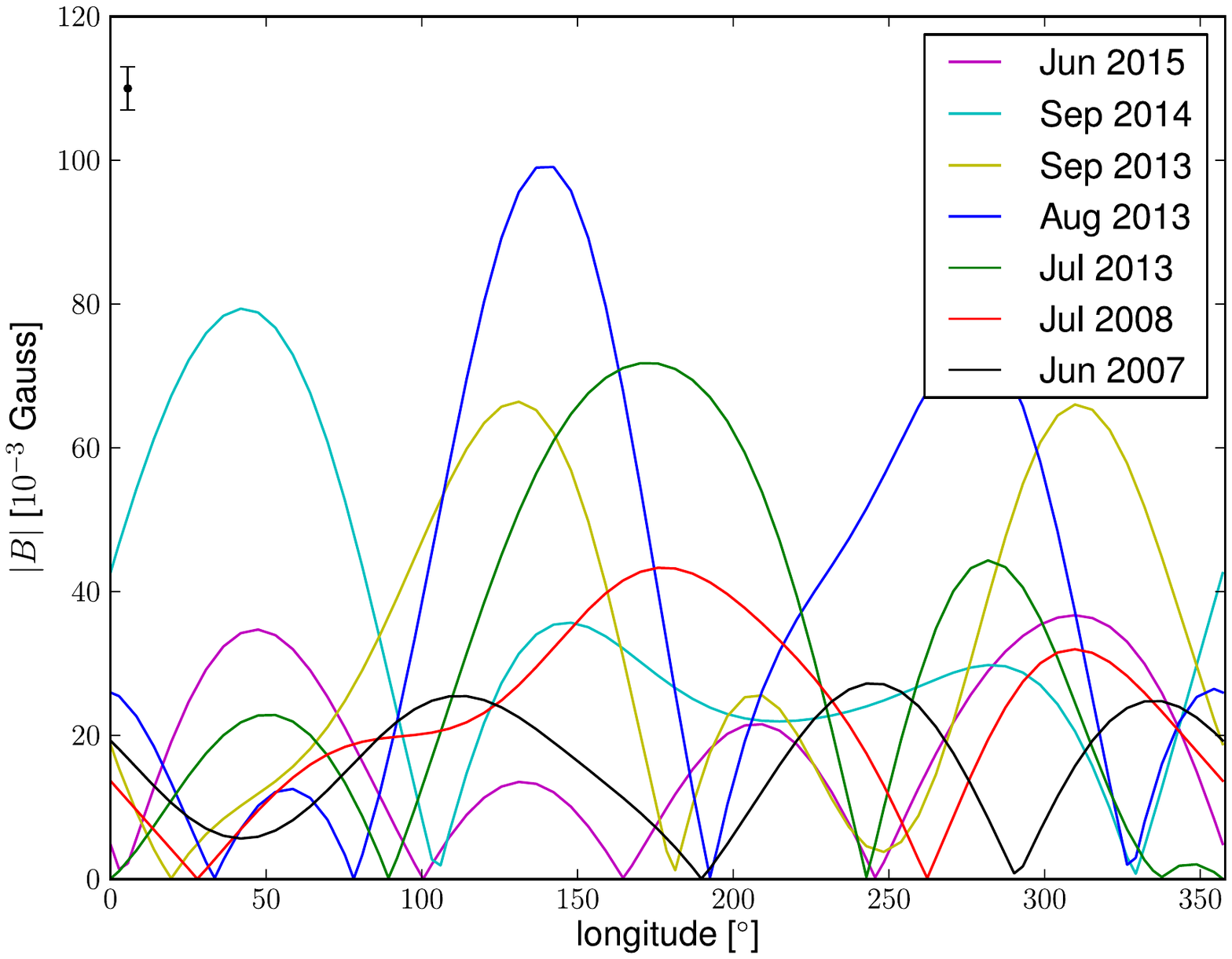}}
\caption{The stellar magnetic field value at the position of the planetary orbit (at 0.031 au). Different colours represent different epochs of observation. This plot shows that the planet is in a non-homogeneous environment, and that this environment varies from one epoch to the other. The error bar shown here is a mean error bar value per orbit, valid for all epochs (see text for more details).}

\label{fig:Bplanet}
\end{figure*}

\section{Discussion: Magnetic Field Evolution}\label{sec:discussion}
In this Section, we take advantage of previous spectropolarimetric observations of HD189733 to investigate the evolution of its magnetic field (intensity and topology) over a 9 year-timespan. The first reconstructed magnetic maps of HD189733, based on ESPaDOnS/CFHT data of June and August 2006, were presented by \citet{moutou07}. Later on, and with the aim of detecting star-planet interaction signatures, \citet{2010MNRAS.406..409F} presented two additional reconstructed magnetic maps of HD189733, based on spectropolarimetric observations of June 2007 and July 2008, as well as an update of the  reconstructed map of 2006, merging June and August data as one dataset.

The data presented in this Paper (5 observing epochs spanning two years) shows that the field of HD~189733 can evolve over a few stellar rotations. Combining datasets spanning more than two stellar rotations systematically reduces the quality of the fit. \cite{2010MNRAS.406..409F} have merged data obtained in 2006, spread over 5 stellar rotations. We revisited the summer 2006 data and found that the map was overfitted. We adopt the results of \cite{moutou07} in this paper. 

Due to the 5-year gap in the observations, we can not investigate the presence of cyclic variations in the stellar magnetic field, in a similar way as those reported for the planet-hosting star $\tau$~Boo \citep{2008MNRAS.385.1179D,2009MNRAS.398.1383F, 2013MNRAS.435.1451F}. Nevertheless, an evolution in the stellar magnetic field intensity and topology can be seen. Variations in both the axisymmetric contribution to the poloidal field and the toroidal contribution to the total field are observed during this time span (see Fig \ref{fig:field_evol}).

For all epochs (apart from June 2006), the toroidal component dominates over the poloidal one. \cite{2008MNRAS.388...80P} suggest, studying a sample of solar-like stars, that the toroidal energy dominates over the poloidal one for stars with rotation periods less than $\sim12$~days. HD~189733, having an equatorial rotation period of 12 days, does not contradict their findings. On the other hand, \cite{donati09} suggest that stars with Rossby number (Ro)  $< 1$ develop toroidal fields. HD~189733 has a $Ro = 0.403$ \citep{2014MNRAS.441.2361V}. HD~189733 field's geometry is therefore compatible with the geometries observed for stars with similar masses and Ro numbers. The fraction of axisymmetric field is almost always less than 50\%.

\begin{table}
\caption{Magnetic field characteristics of HD189733 for different epochs. The columns are: the epoch of the observations, the mean magnetic field at the surface of the star, the percentage of the toroidal energy relative to the total one, the percentage of the energy contained in the axisymmetric modes of the poloidal component relative to the poloidal energy, the percentage contribution of the dipolar, quadrupolar and octupolar components to the poloidal energy, and the mean stellar field at the position of the planetary orbit (see Section \ref{sec:extrapol}). Results of 2006 are from \protect\cite{moutou07} and the results of 2007 and 2008 are from \protect\cite{2010MNRAS.406..409F}. Error bars are calculated as in \protect\cite{2016MNRAS.459.4325M}, i.e. by varying the input parameters within their error bars. The error bar on $B_{orbit}$ is of the order of 3 mG (see text for more details).}

\label{tab:prop}
\begin{tabular}{cccccccc}
\hline
Epoch & $B_{mean}$& $E_{\rm tor}$& $E_{\rm axi}$ & $E_{\rm l=1}$ &$E_{\rm l=2}$ &$E_{\rm l=3}$ & $B_{orbit}$\\
& (G)&\%&\%&\%&\%&\% & (mG)\\

\hline
7/2015 & $37^{+2}_{-2}$ & $85^{+2}_{-2}$ & $9^{+2}_{-2}$ &$33^{+5}_{-2}$ & $32^{+2}_{-4}$ & $10^{+10}_{-1}$ & 18 \\

9/2014 & $32^{+2}_{-4}$ & $78^{+3}_{-5}$ & $ 10^{+2}_{-7} $ & $21^{10}_{-6} $& $35^{+6}_{-2}$ & $16^{+1}_{-6}$ & 33\\ 

9/2013 & $42^{+2}_{-4}$ & $59^{+1}_{-4}$ & $2^{+2}_{-1}$ & $4^{+4}_{-1}$ & $3^{+2}_{-1}$ &$ 49^{+2}_{-2}$ & 31 \\

8/2013 & $41^{+2}_{-5}$ & $ 50^{+5}_{-5} $ & $2_{-1} $ &  $10^{+5}_{-2}$ & $20^{+5}_{-1}$&$32^{+3}_{-2}$&  39\\

6/2013& $36^{+4}_{-3}$ & $61^{+4}_{-3}$ & $38^{+1}_{-2}$ & $21^{+1}_{-1}$ & $37^{+3}_{-3}$ & $17^{+5}_{-1}$& 30 \\ 

7/2008 & $36^{+1}_{-3}$ &  $77^{+3}_{-3}$ & $17^{+2}_{-7}$ & $30^{+4}_{-5}$ & $26^{+2}_{-8}$ & $12_{-2}$&23\\  

6/2007 & $22_{-3}$ &$57^{+8}$ & $26_{-5}$ & $7^{+2}$       & $33^{+7} $        & $30^{+2}_{-1}$        &16\\

8/2006 & $20$ & $60$ & $10$ & 35 &20 &13 \\
6/2006 & $18$ & $ 35 $ & $52$ & 50& 36&12 \\

\hline
\end{tabular}
\label{table:stat}
\end{table}

In addition, it is also interesting to compare the evolution of the stellar magnetic field within the 2013 data sets, that are separated by just $9$ rotation periods. We note that little variation was seen for the toroidal component. The percentage of the contributors (dipole, quadrupole, octupole) to the poloidal field, on the other hand, has changed. The main contributor to the field (i.e. azimuthal component) does not change polarity. The radial component evolved, with negative and positive magnetic features appearing at the surface.

\begin{figure*}
\center{
\hbox{\includegraphics[scale=0.3]{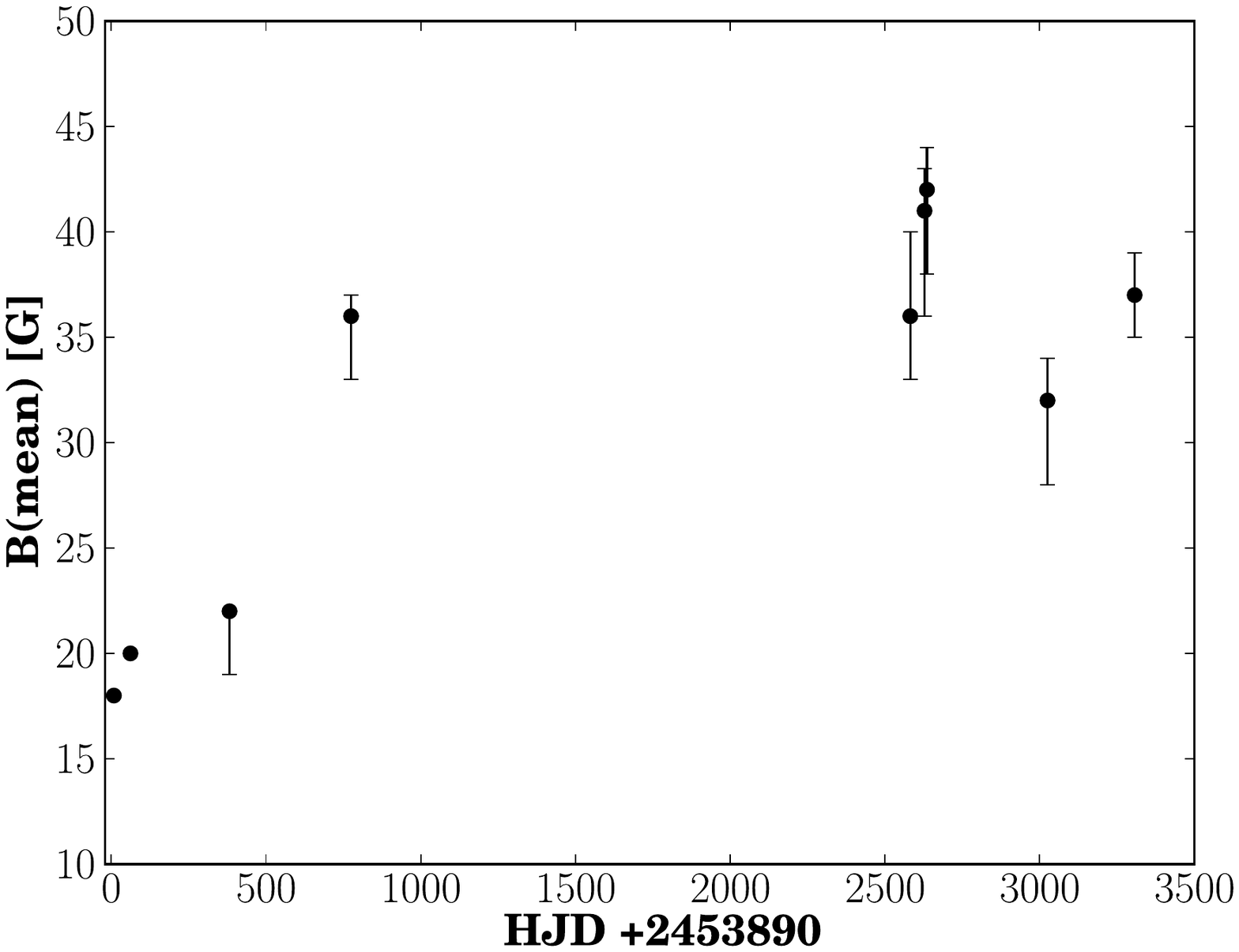}\includegraphics[scale=0.3]{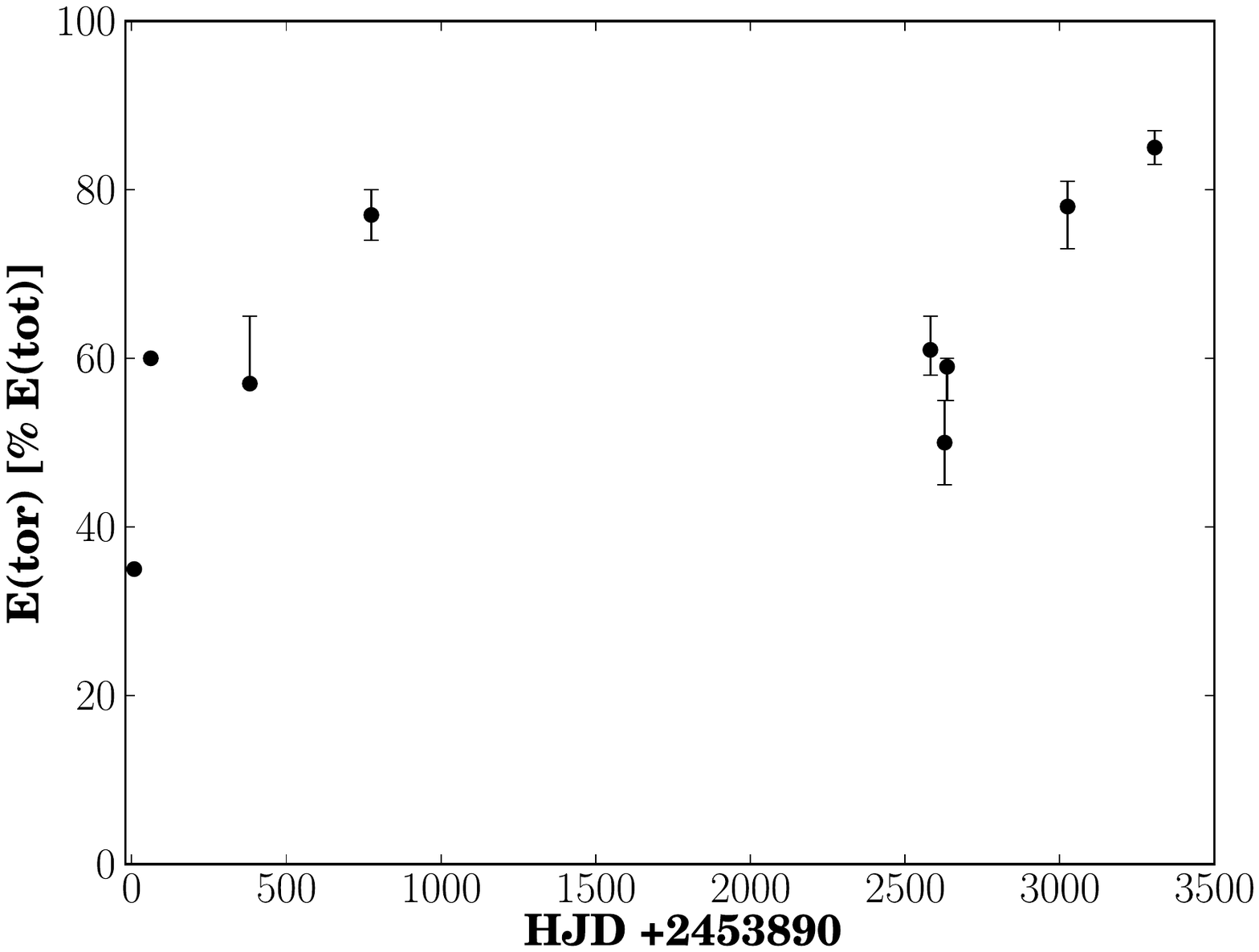}\includegraphics[scale=0.3]{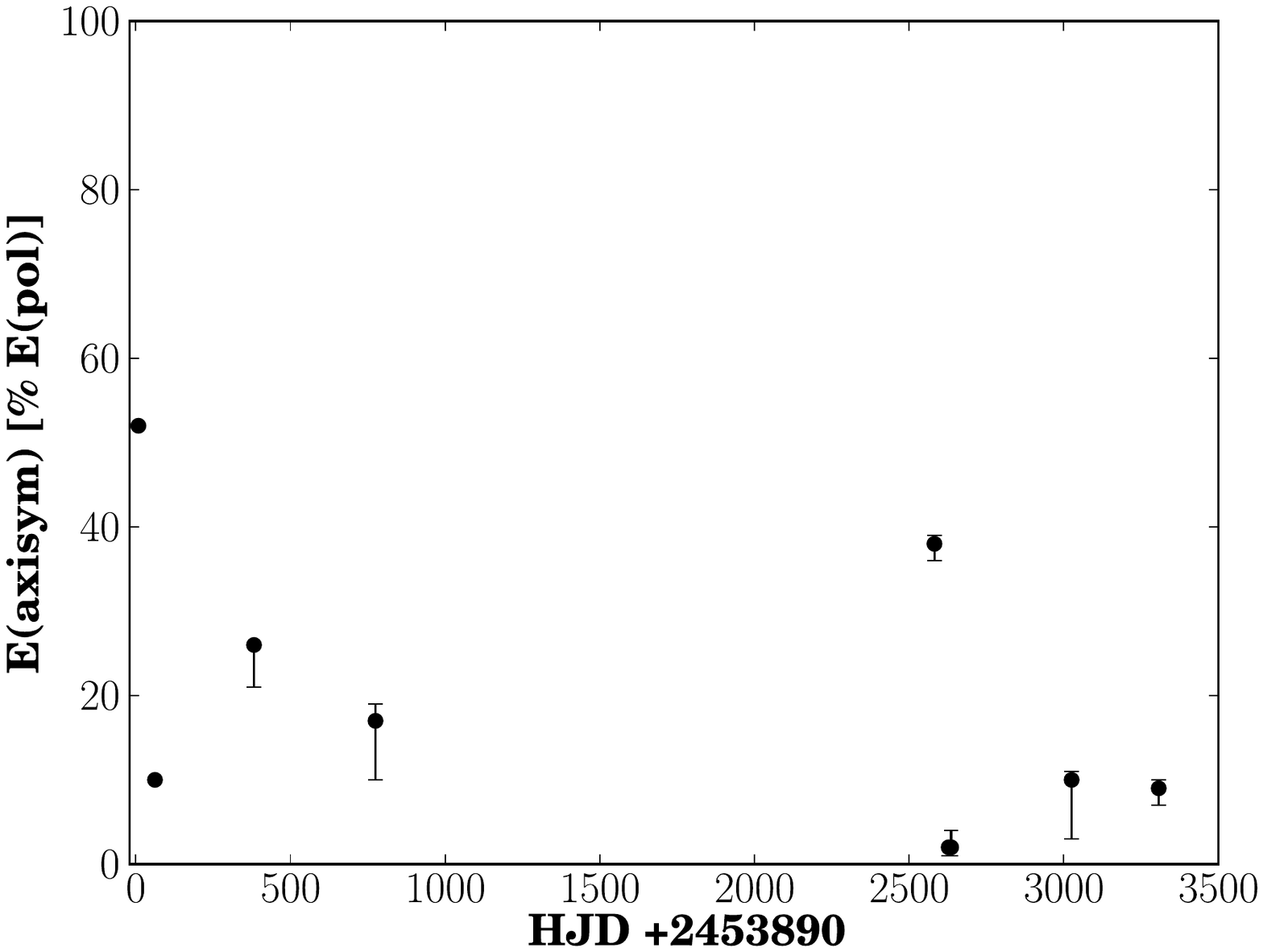}}}
\caption{Magnetic field evolution of HD189733. From left to right: the square-root of the total magnetic energy, the toroidal energy relative to the total one, and the energy in the axisymmetric modes of the poloidal component relative to the energy of the poloidal component. Error bars are calculated as stated in the text.}
\label{fig:field_evol}
\end{figure*}

\section{Conclusions and Future plans}\label{sec:conclusions}

This paper is part of the MOVES collaboration (Multiwavelength Observations of an eVaporating Exoplanet and its Star), which aims to characterise comprehensively the complex environment of the exoplanet HD189733b. Orbiting a bright and active K dwarf at short distance, this transiting hot-Jupiter has been subjected to many stellar and planetary atmosphere studies. The main objectives of MOVES are to probe the different regions of the extended planetary atmosphere, its interactions with the host-star, and their temporal variability. The wider set of multi-wavelength observations (X-ray with Swift and XMM-Newton, UV spectroscopy with HST and XMM-Newton, and radio observations with LOFAR) were taken contemporaneously with the magnetic field mapping presented here.

In this first paper, we presented a detailed spectropolarimetric study of HD189733 and studied the evolution of its magnetism. Stellar magnetism is an important ingredient in stellar evolution, and also has important effects on planets surrounding these stars. The star was observed at five epochs (Jul~2013, Aug~2013, Sept~2013, Sept~2014 and Jun~2015), during which we also collected X-ray and UV observations (Wheatley et al, in prep). Using Zeeman Doppler Imaging, we reconstructed the magnetic maps of the star. With a strength up to 45~G, the magnetic field is dominated by the toroidal component at the five epochs. The toroidal component is mainly axisymmetric during all observing epochs. In contrast, the poloidal component is mainly non-axisymmetric. We will continue monitoring this system to study the magnetic evolution on time-scales longer than 2 years and look for a potential magnetic cycle. These reconstructed magnetic maps are crucial for analysing multi-wavelength observations. They allow us, modelling the stellar wind in the corona and at the planetary orbit, to reconstruct the X-ray emission and irradiation of the planet, as well as the spatial distribution of X-ray that will be absorbed by the extended atmosphere of the exoplanet.

\section*{Acknowledgments}
The authors thank an anonymous referee for their useful comments. This work is based on observations obtained with ESPaDOnS at the Canada-France-Hawaii Telescope (CFHT) and with NARVAL at the T\'elescope Bernard Lyot (TBL). CFHT/ESPaDOnS are operated by the National Research Council of Canada, the Institut National des Sciences de l'Univers of the Centre National de la Recherche Scientifique (INSU/CNRS) of France, and the University of Hawaii, while TBL/NARVAL are operated by INSU/CNRS. We thank the CFHT and TBL staff for their help during the observations, and in particular R. Cabanac and P. Petit. We also thank J.-F. Donati and E. H\'ebrard for useful comments on the data analysis. RF acknowledges financial support by WOW from INAF through the \textit{Progetti Premiali} funding scheme of the Italian Ministry of Education, University, and Research. VB and AL acknowledge the support of the French Agence Nationale de la Recherche (ANR), under program ANR-12-BS05-0012 `Exo-Atmos'. Part of VB work has been carried out in the frame of the National Centre for Competence in Research ``PlanetS'' supported by the Swiss National Science Foundation (SNSF). V.B. also acknowledges the financial support of the SNSF. AAV acknowledges partial support from an Ambizione Fellowship of the Swiss National Science Foundation. ChH highlights financial support of the European Community under the FP7 by an ERC starting grant number 257431. P.W. is supported by a STFC consolidated grant (ST/L000733/7)
\bibliographystyle{mn2e}

\def\aj{{AJ}}                   
\def\araap{{ARA\&A}}             
\def\apj{{ApJ}}                 
\def\apjl{{ApJ}}                
\def\apjs{{ApJS}}               
\def\ao{{Appl.~Opt.}}           
\def\apss{{Ap\&SS}}             
\def\aap{{A\&A}}                
\def\aapr{{A\&A~Rev.}}          
\def\aaps{{A\&AS}}              
\def\azh{{AZh}}                 
\def\baas{{BAAS}}               
\def\jrasc{{JRASC}}             
\def\memras{{MmRAS}}            
\def\mnras{{MNRAS}}             
\def\pra{{Phys.~Rev.~A}}        
\def\prb{{Phys.~Rev.~B}}        
\def\prc{{Phys.~Rev.~C}}        
\def\prd{{Phys.~Rev.~D}}        
\def\pre{{Phys.~Rev.~E}}        
\def\prl{{Phys.~Rev.~Lett.}}    
\def\pasp{{PASP}}               
\def\pasj{{PASJ}}               
\def\qjras{{QJRAS}}             
\def\skytel{{S\&T}}             
\def\solphys{{Sol.~Phys.}}      
\def\sovast{{Soviet~Ast.}}      
\def\ssr{{Space~Sci.~Rev.}}     
\def\zap{{ZAp}}                 
\def\nat{{Nature}}              
\def\iaucirc{{IAU~Circ.}}       
\def\aplett{{Astrophys.~Lett.}} 
\def\apspr{{Astrophys.~Space~Phys.~Res.}}   
\def\bain{{Bull.~Astron.~Inst.~Netherlands}}    
\def\fcp{{Fund.~Cosmic~Phys.}}  
\def\gca{{Geochim.~Cosmochim.~Acta}}        
\def\grl{{Geophys.~Res.~Lett.}} 
\def\jcp{{J.~Chem.~Phys.}}      
\def\jgr{{J.~Geophys.~Res.}}    
\def\jqsrt{{J.~Quant.~Spec.~Radiat.~Transf.}}   
\def\memsai{{Mem.~Soc.~Astron.~Italiana}}   
\def\nphysa{{Nucl.~Phys.~A}}    
\def\physrep{{Phys.~Rep.}}      
\def\physscr{{Phys.~Scr}}       
\def\planss{{Planet.~Space~Sci.}}           
\def\procspie{{Proc.~SPIE}}     
\def\icarus{{Icarus}}     

\bibliography{biblio}

\end{document}